\documentclass[11pt, a4paper]{appolb}
\usepackage{graphicx}
\usepackage{xspace}
\usepackage{amsmath,amssymb}
\usepackage{wasysym}
\usepackage{slashed}
\usepackage[usenames,dvipsnames]{xcolor}
\usepackage[separate-uncertainty=true]{siunitx}
\usepackage{tabularx}
\usepackage{soul}
\usepackage{color}

\newcommand{\TeV}{{\ensuremath\rm TeV}\xspace}
\newcommand{\GeV}{{\ensuremath\rm GeV}\xspace}
\newcommand{\fb}{{\ensuremath\rm fb}\xspace}
\newcommand{\pb}{{\ensuremath\rm pb}\xspace}
\newcommand{\ab}{{\ensuremath\rm ab}\xspace}
\newcommand{\eqn}{equation}
\newcommand{\lam}{\lambda}
\newcommand{\Ztwo}{\ensuremath{{\mathbb{Z}_2}}\xspace}
\newcommand{\eqdot}{\,.}
\newcommand{\eqcomma}{\,,}

\newcommand{\HSv}[1]{\texttt{HiggsSignals-#1}}
\newcommand{\HBv}[1]{\texttt{HiggsBounds-#1}}
\newcommand{\HS}{\texttt{HiggsSignals}}
\newcommand{\HB}{\texttt{HiggsBounds}}
\newcommand{\lb}{\left(}
\newcommand{\rb}{\right)}

\begin{document}
%\preprint{RBI-ThPhys-2021-14}
\title{Extended scalar sectors at future colliders%
\thanks{Presented by Tania Robens at XXVII Cracow EPIPHANY Conference of Future of particle physics}%
}
\author{Tania Robens\\
\vspace{-3mm}
\address{Theoretical Physics Division, Rudjer Boskovic Institute,
10002 Zagreb, Croatia}
\\
\vspace{4mm}
Jan Kalinowski, Aleksander Filip \.Zarnecki\\
\vspace{-3mm}
\address{Faculty of Physics, University of Warsaw, 
  ul.~Pasteura 5, 02--093 Warsaw, Poland}
\\
\vspace{4mm}
Andreas Papaefstathiou\\
\vspace{-3mm}
\address{Department of Physics, Kennesaw State University, Kennesaw, GA 30144, USA}
}

\maketitle
\begin{abstract}
After the discovery of the Higgs boson in 2012, particle physics has entered an exciting era. An important question is whether the Standard Model of particle physics correctly describes the scalar sector realized by nature, or whether it is part of a more extended model, featuring additional particle content. A prime way to test this is to probe models with extended scalar sectors at future collider facilities. We here discuss such models in the context of high-luminosity LHC, a possible proton-proton collider with 27 and 100 \TeV center-of-mass energy, as well as future lepton colliders with {various} center-of-mass energies.

\end{abstract}
RBI-ThPhys-2021-14
  
\section{Introduction}
After the discovery of a scalar which complies with the properties of the Higgs boson {$h$} of the Standard Model {(SM)} {\cite{Aad:2012tfa,Chatrchyan:2012ufa}}, particle physics has entered an exciting era. One important question is whether the particle discovered by the LHC experiments indeed corresponds to the Higgs boson predicted by the Standard Model, or whether it is part of a model featuring an extended scalar sector. So far, both theoretical as experimental uncertainties allow for both options, although in general the parameter space of new physics models becomes increasingly constrained by
both direct searches as well as indirect probes, as e.g. the 125 \GeV~{scalar} coupling strength or electroweak precision observables {(see e.g. \cite{atlpub, cmspub} for recent results)}.

In this work, we will discuss the discovery prospects of models with extended scalar sectors at future collider facilities. We will concentrate on two different scenarios: {\sl (a)} The Inert Doublet Model, a two-Higgs-doublet model that features an exact $\mathbb{Z}_2$ symmetry. This model introduces 5 additional particles in the so-called dark sector and provides a dark matter candidate. {\sl (b)} The two-real singlet extension of the Standard Model, where the scalar sector is enhanced by two real scalar singlets. In this model, {an} exact $\mathbb{Z}_2\,\otimes\,\mathbb{Z}_2'$ symmetry {is assumed, which} is {successively} broken by the vacuum expectation values { (vevs)} of the two additional scalar fields. {This} induces mixing between all CP-even scalar states. This model features a plethora of new physics channels which so far have not been investigated by the LHC experiments. We will briefly discuss these and subsequently focus on the discovery prospects for ${h}{h}{h}$ production within this model at LHC Run III as well as High-Luminosity LHC.

The two models described above will be covered in the subsequent sections of this manuscript.
Section \ref{sec:idm} will discuss the Inert Doublet Model, current status as well as possible future collider prospects. In section \ref{sec:2rs}, on the other hand, we will comment on the two-real-singlet extension, with a special focus on discovery prospects for the $hhh$ final state at the HL-LHC. We will conclude in section \ref{sec:concl}.

\section{The Inert Doublet Model}\label{sec:idm}
\subsection{The model}
The Inert Doublet Model (IDM) \cite{Deshpande:1977rw,Cao:2007rm,Barbieri:2006dq} is an intriguing new physics model that enhances the SM scalar sector by an additional $SU(2)\,\times\,U(1)$ gauge doublet $\phi_D$. Furthermore, it introduces a discrete $\mathbb{Z}_2$ symmetry with the following transformation properties
\begin{equation}\label{eq:symm}
\phi_S\to \phi_S, \,\, \phi_D \to - \phi_D, \,\,
\text{SM} \to \text{SM}.
\end{equation}
In this model the symmetry remains exact. This has important consequences: {\sl (a)} the additional doublet does not acquire a vacuum expectation value (vev) and {\sl (b)} it does not couple to fermions. Therefore, electroweak symmetry breaking proceeds as in the SM. Furthermore, the above symmetry insures that the lightest particle of the so-called dark doublet $\phi_D$ is stable and renders a dark matter candidate.

The potential of the model is given by
\begin{equation}\begin{array}{c}
V=-\frac{1}{2}\left[m_{11}^2(\phi_S^\dagger\phi_S)\!+\! m_{22}^2(\phi_D^\dagger\phi_D)\right]+
\frac{\lambda_1}{2}(\phi_S^\dagger\phi_S)^2\! 
+\!\frac{\lambda_2}{2}(\phi_D^\dagger\phi_D)^2\\[2mm]+\!\lambda_3(\phi_S^\dagger\phi_S)(\phi_D^\dagger\phi_D)\!
\!+\!\lambda_4(\phi_S^\dagger\phi_D)(\phi_D^\dagger\phi_S) +\frac{\lambda_5}{2}\left[(\phi_S^\dagger\phi_D)^2\!
+\!(\phi_D^\dagger\phi_S)^2\right].
\end{array}\label{pot}\end{equation}

After electroweak symmetry breaking, the model features 7 free parameters. We here chose these in the so-called physical basis {\cite{Ilnicka:2015jba}}
\begin{\eqn}\label{eq:physbas}
v,M_h,M_H, M_A, M_{H^{\pm}}, \lam_2, \lam_{345},
\end{\eqn}
where we use $\lam_{345}\,\equiv\,\lam_3+\lam_4+\lam_5$ throughout this work.
The vev $v$ as well as $M_h\,\sim\,125\,\GeV$ are fixed by experimental measurements, leading to a total number of 5 free parameters. We here choose $H$ as the dark matter candidate, which implies $M_{A,\,H^\pm}\,\geq\,M_H$. \footnote{Note that the new scalars in the IDM do not have CP quantum numbers, as they do not couple to fermions. In the subsequent discussion, we can replace $H\,\longleftrightarrow\,A$ if we simultaneously use $\lam_5\,\longleftrightarrow\,-\lam_5$. All phenomenological considerations are identical for these cases.}

The model is subject to a large number of theoretical and experimental constraints. These have been discussed at length e.g. in \cite{Ilnicka:2015jba,Ilnicka:2018def,Dercks:2018wch,Kalinowski:2018ylg,Kalinowski:2020rmb} and will therefore not be repeated here. In the scan for the allowed parameter ranges, we make use of the publicly available tools \texttt{2HDMC} \cite{Eriksson:2009ws},  \HBv5.9.0 \cite{Bechtle:2008jh, Bechtle:2011sb, Bechtle:2013wla,Bechtle:2015pma,Bechtle:2020pkv}, \HSv2.6.0 \cite{Bechtle:2013xfa,Bechtle:2020uwn}, as well as \texttt{micrOMEGAs$\_$5.2.4} \cite{Belanger:2020gnr}. Cross sections are calculated using  {\texttt{Madgraph5}} \cite{Alwall:2011uj} with a UFO input file from \cite{Goudelis:2013uca}\footnote{\label{foot:ufo} Note the official version available at \cite{ufo_idm} exhibits a wrong CKM structure, leading to false results for processes involving electroweak gauge bosons radiated off quark lines. In our implementation, we corrected for this. Our implementation corresponds to the expressions available from \cite{Zyla:2020zbs}.}. We compare to experimental values from GFitter \cite{gfitter,Haller:2018nnx}, as well as results from the Planck \cite{Aghanim:2018eyx} and XENON1T \cite{Aprile:2018dbl} experiments. Direct collider searches as well as agreement with the 125 \GeV coupling strength measurements are implemented via \HB~ and \HS, where we additionally compare to the total width upper limit  \cite{Sirunyan:2019twz} and invisible branching ratio \cite{ATLAS-CONF-2020-052} of $h$. Recast results from a LEP-SUSY search \cite{Lundstrom:2008ai} were also included.{We} refer {the reader} to the above references for more details.

\subsection{Current Status}
\begin{figure}[htb]
\centerline{%
\includegraphics[width=0.48\textwidth]{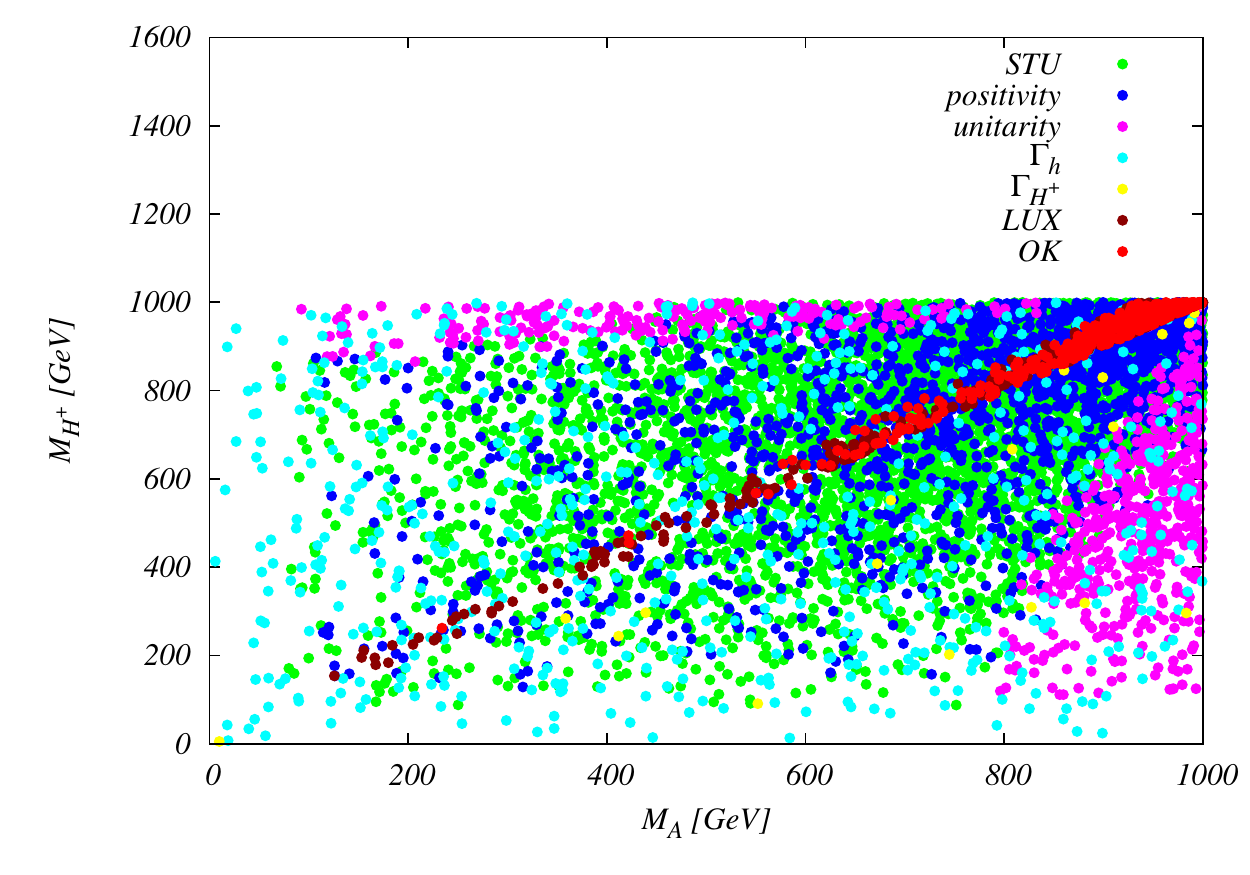}
\includegraphics[width=0.48\textwidth]{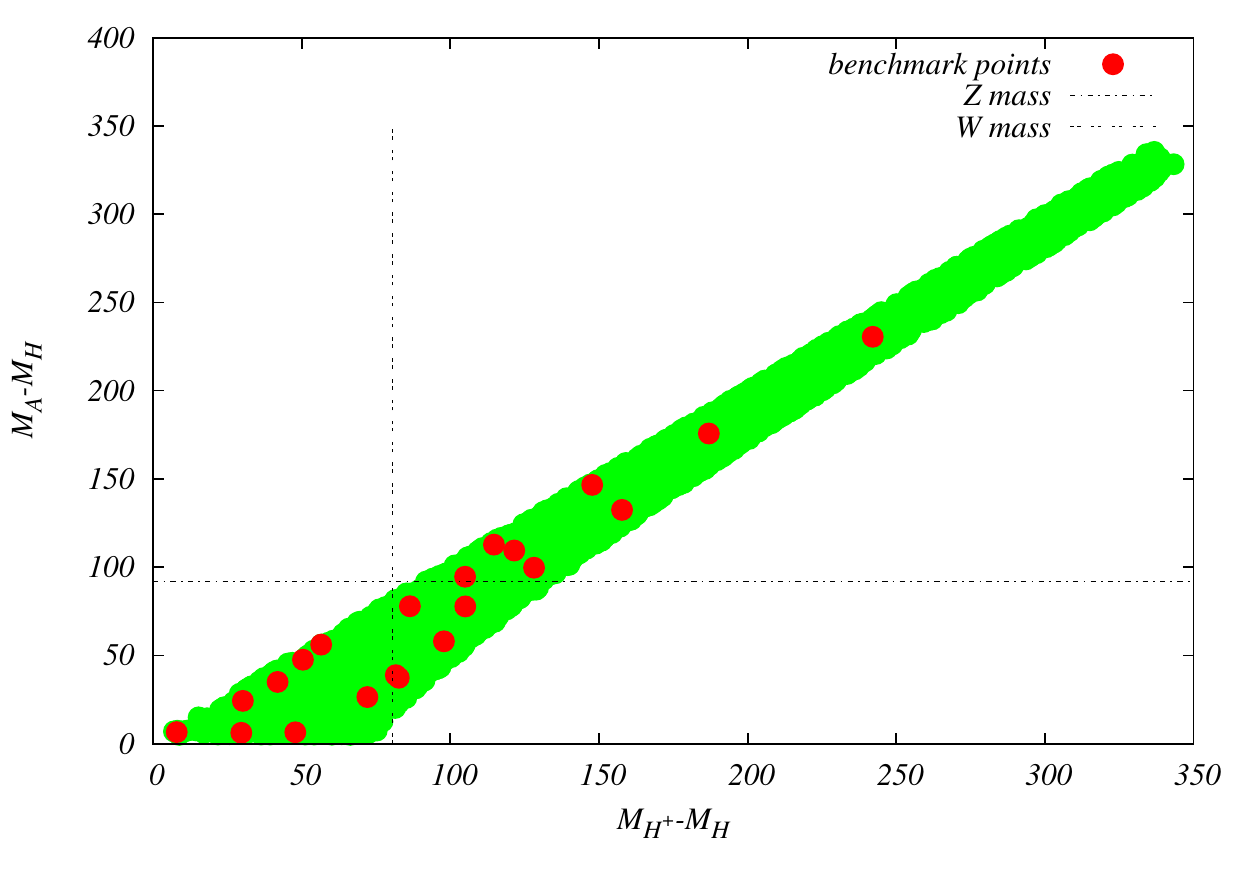}}
\caption{Masses are requested to be quite degenerate after all constraints have been taken into account. {\sl Left:} In the $\lb M_A,\,M_{H^\pm} \rb$ plane (taken from \cite{Ilnicka:2015jba}). {\sl Right:} In the {$\lb M_{H^\pm}-M_H,\,M_A-M_H \rb$} plane (taken from \cite{Kalinowski:2018ylg}).}
\label{fig:massesidm}
\end{figure}

\begin{figure}[htb]
\centerline{%
\includegraphics[width=0.48\textwidth]{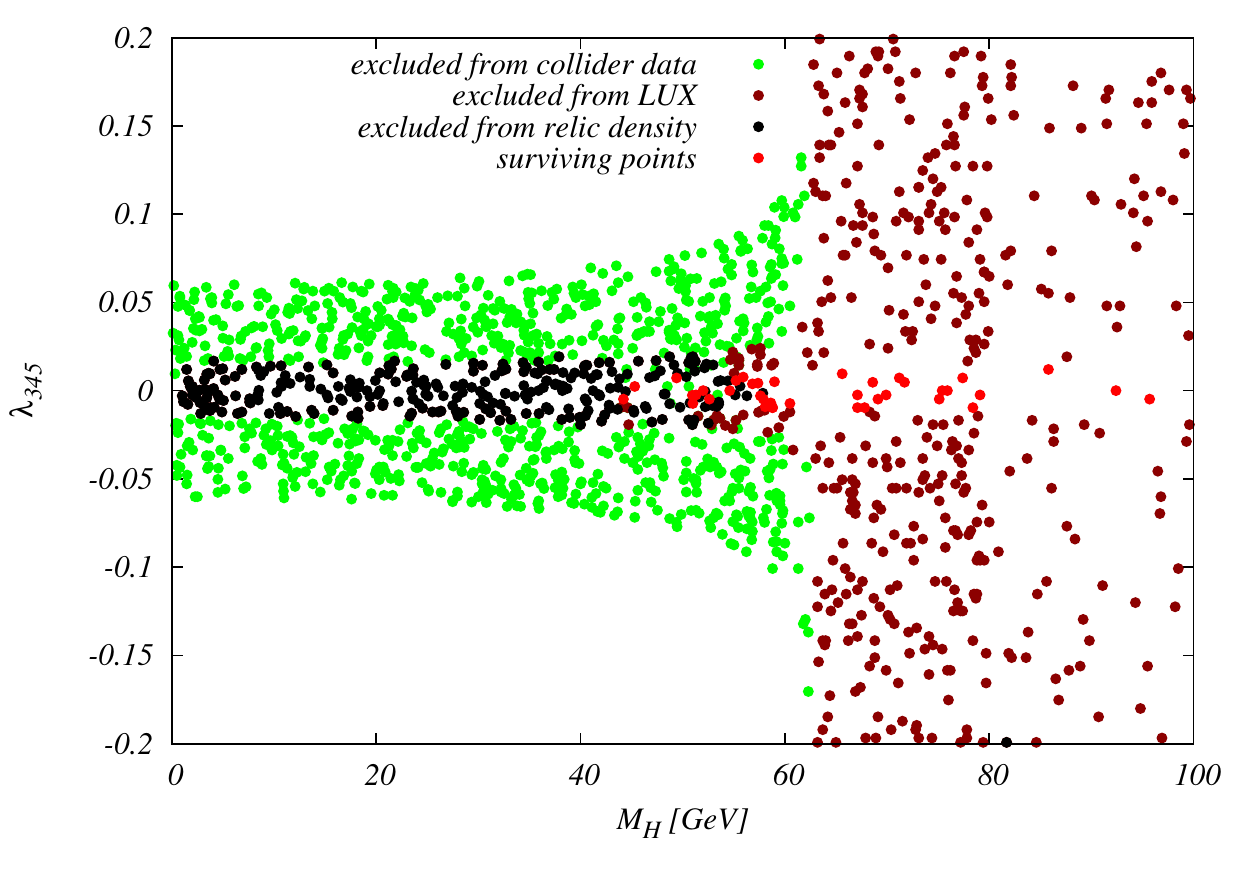}
\includegraphics[width=0.48\textwidth]{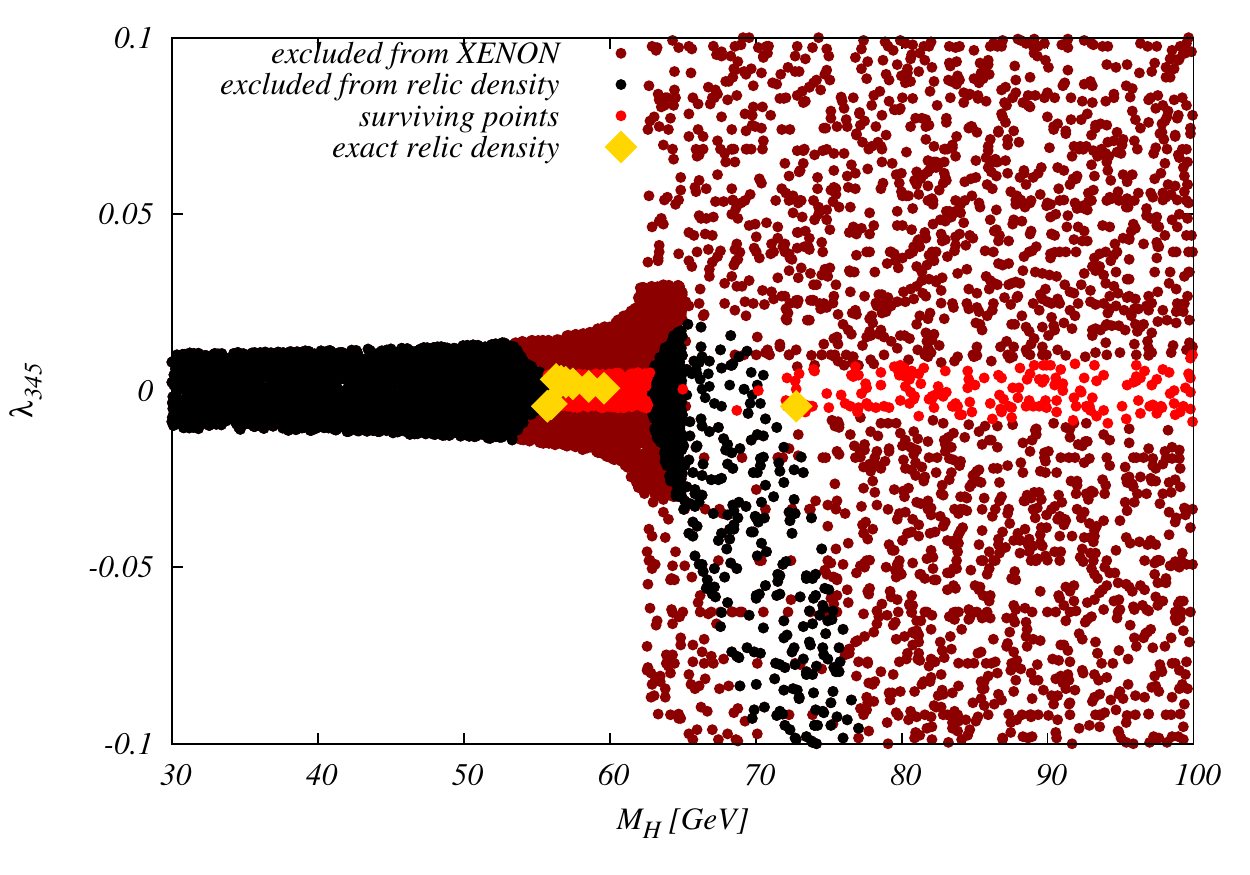}}
\caption{Interplay of signal strength and relic density constraints in the $\lb M_H,\,\lam_{345}\rb$ plane. {\sl Left:} Using LUX constraints \cite{Akerib:2013tjd}, bounds labelled "excluded from collider data" have been tested using \HB~and \HS~(taken from \cite{Ilnicka:2015jba}). {\sl Right:} Using XENON1T results, with golden points labelling those points that produce exact relic density (taken from \cite{Ilnicka:2018def}).}
\label{fig:lowmh}
\end{figure}

The experimental and theoretical constraints lead to a large reduction of the allowed parameter space of the model; in particular, the masses are usually constrained to be quite degenerate, as can be seen {from} figure \ref{fig:massesidm}. This is due to an interplay of electroweak constraints as well as theoretical requirements on the potential. 
A particularly interesting scenario is the case when $M_H\,\leq\,M_h/2$, which opens up the $h\,\rightarrow\,\text{inv{isible}}$ channel. In such a scenario, there is an interesting interplay between bounds from signal strength measurements, that require $|\lam_{345}|$ to be rather small $\lesssim\,0.3$, and bounds from dark matter relic density, where too low values of that parameter lead to small annihilation cross sections and therefore too large relic density values. The effects of this are shown in figure \ref{fig:lowmh}. In \cite{Ilnicka:2015jba}, it was found that this in general leads to a lower bound of $M_H\,\sim\,50\,\GeV$, although exceptions to this rule were presented in \cite{Kalinowski:2020rmb}. 
\subsection{Discovery prospects at CLIC}
So far, no publicly available search exists that investigates the IDM parameter space with actual collider data. In \cite{Kalinowski:2018kdn,deBlas:2018mhx}, however, the discovery potential of CLIC was investigated for several benchmark points proposed in \cite{Kalinowski:2018ylg}, for varying center-of-mass energies up to $3\,\TeV$. We investigated both $AH$ and $H^+ H^-$ production with $A\,\rightarrow\,Z\,H$ and $H^\pm\,\rightarrow\,W^\pm H$, where the electroweak gauge bosons subsequently decay leptonically. Event generation was performed using \texttt{WHizard 2.2.8} \cite{Moretti:2001zz,Kilian:2007gr}, with an interface via \texttt{SARAH} \cite{Staub:2015kfa} and \texttt{SPheno 4.0.3} \cite{Porod:2003um,Porod:2011nf} for model implementation. CLIC energy spectra \cite{Linssen:2012hp} were also taken into account.

\begin{figure}[htb]
\begin{center}%{%
\includegraphics[width=0.48\textwidth]{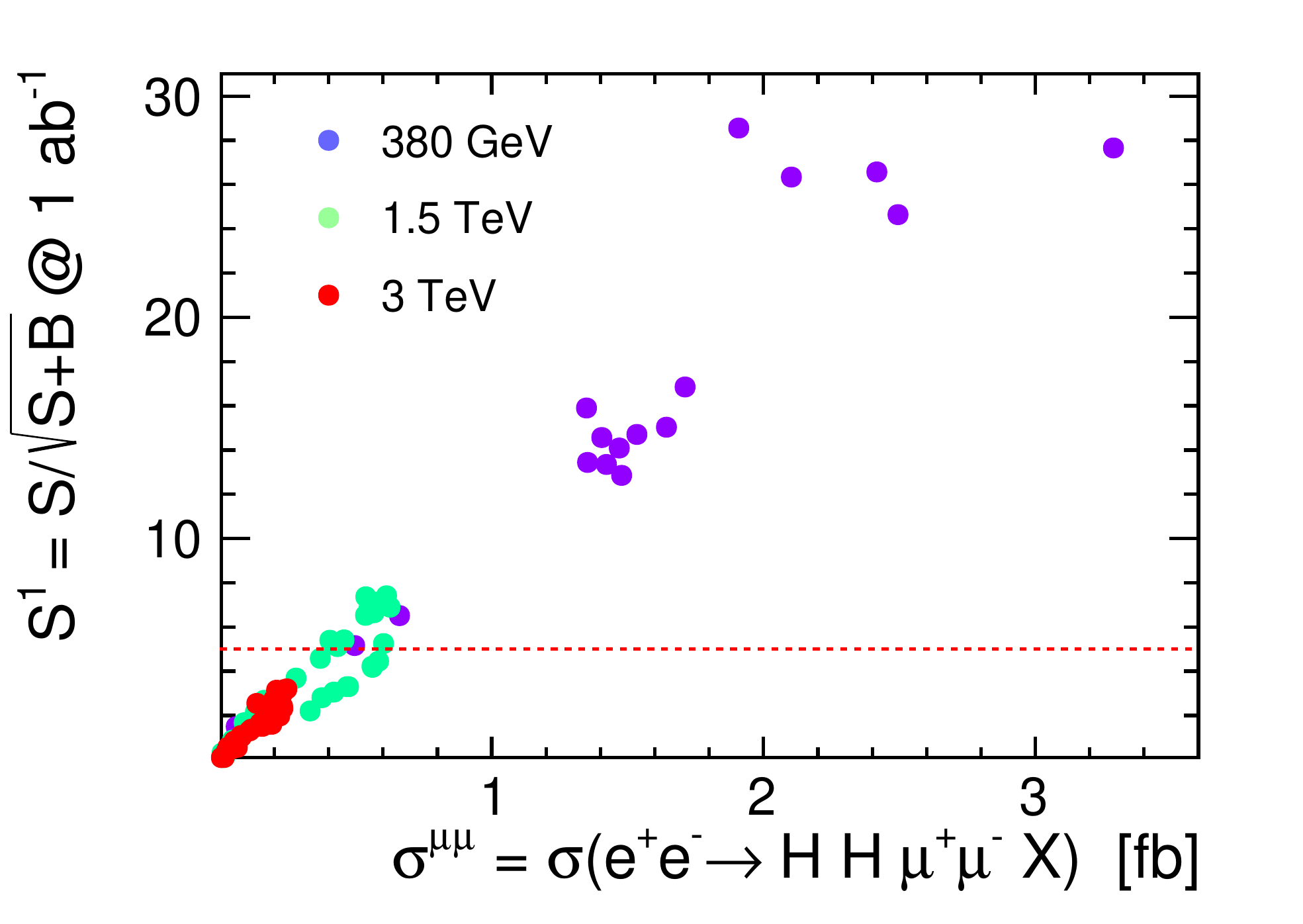}
\includegraphics[width=0.48\textwidth]{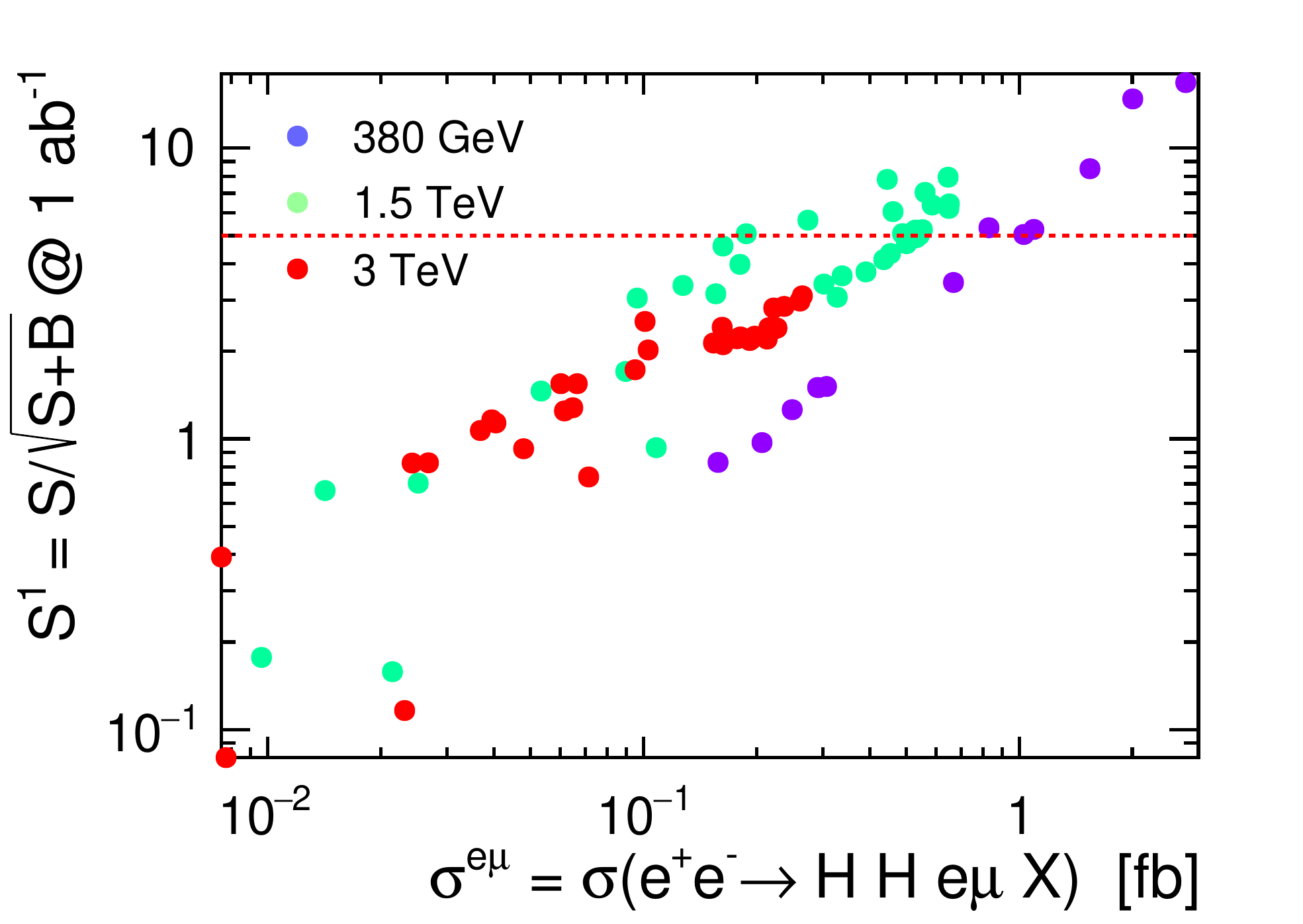}
\includegraphics[width=0.48\textwidth]{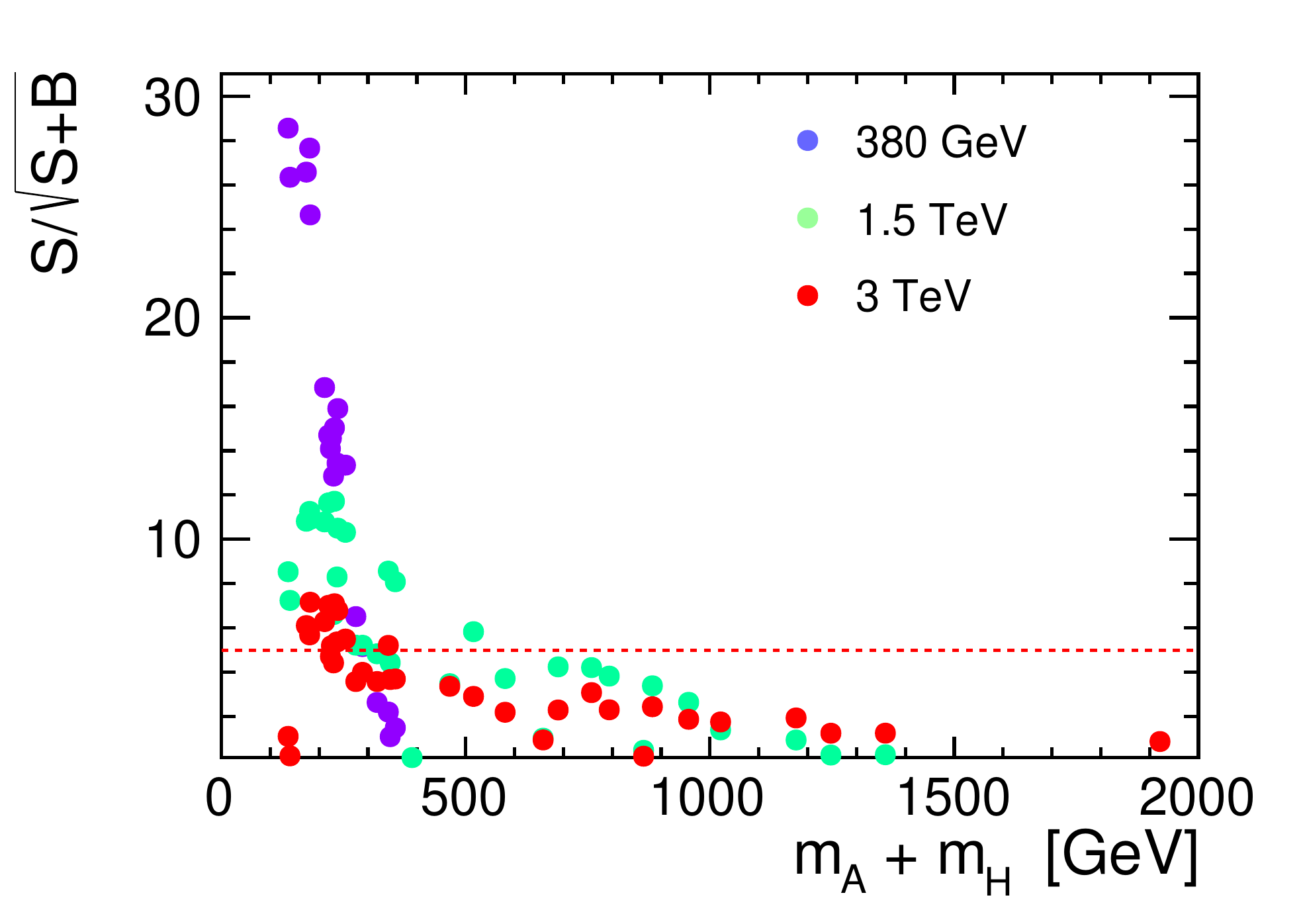}
\includegraphics[width=0.48\textwidth]{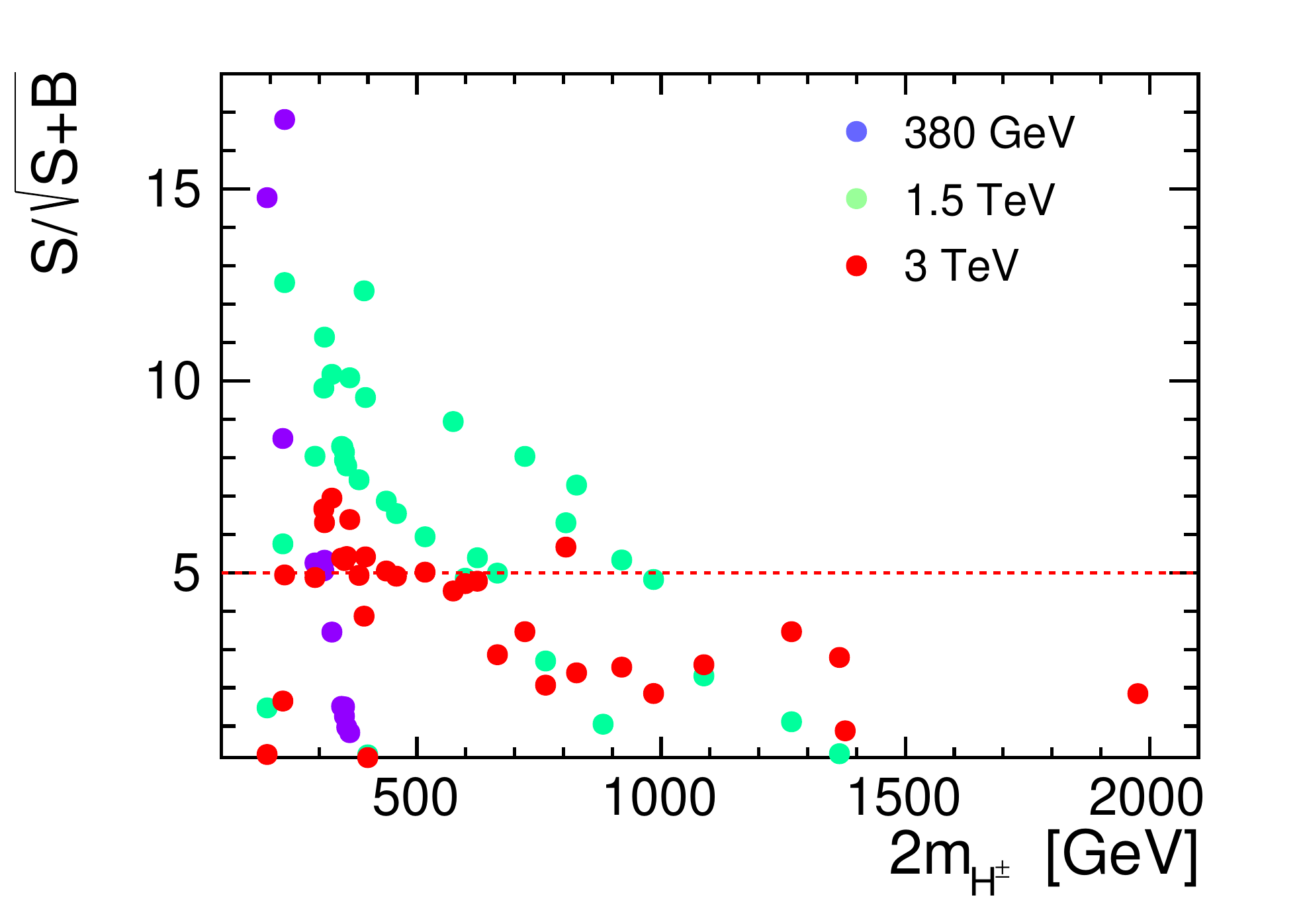}
\end{center}
%}
\caption{Discovery prospects at CLIC for the IDM in $\mu^+\mu^-+\slashed{E}$ {\sl (left)} and $\mu^\pm\,e^\mp+\slashed{E}$ {\sl (right)} final states, as a function of the respective production cross-sections {\sl (top)} and mass sum of the produced particles {\sl (bottom)}. Taken from \cite{Kalinowski:2018kdn}.}
\label{fig:clic}
\end{figure}

For the production modes above, we considered leptonic decays of the electroweak gauge bosons. In particular, the investigated final states were
\begin{\eqn*}
e^+\,e^-\,\rightarrow\,\mu^+\mu^-+\slashed{E},\,e^+\,e^-\,\rightarrow\,\mu^\pm\e^\mp+\slashed{E}
\end{\eqn*}
for $HA$ and $H^+\,H^-$ production, respectively. Note however, that in the event generation we did not specify the intermediate states, which means all processes leading to the above signatures were taken into account, including interference between the contributing diagrams. This includes final states where the missing energy can originate from neutrinos in the final state. \\
Event selection was performed using a set of preselection cuts as well as boosted decision trees, as implemented in the TMVA toolkit \cite{Hocker:2007ht}. Results for the discovery reach of CLIC with varying center of mass energies {are shown} in figure \ref{fig:clic}.  We see that in general, production cross sections $\gtrsim\,0.5\,\fb$ seem to be accessible, where best prospects for the considered benchmark points are given for 380 \GeV or 1.5 \TeV center-of-mass energies. {Similarly}, mass sums up to 1 \TeV seem accessible, where in general the $\mu^\pm\,e^\mp$ channel seems to provide a larger discovery range.
Considering {the $H^+\,H^-$ production with the semi-leptonic final state, i.e. with hadronic decay of one of the $W$ bosons}, increases the {corresponding} mass range to about 2 \TeV \cite{Sokolowska:2019xhe,Zarnecki:2020swm,Zarnecki:2020nnw,Klamka:2728552}.

\subsection{Sensitivity comparison at future colliders}
After a dedicated analysis of the IDM benchmarks in the CLIC environment, an important question is whether other current or future collider options provide similar or better discovery prospects. Therefore, for the benchmarks proposed in \cite{Kalinowski:2018ylg,Kalinowski:2018kdn}, production cross sections for a variety of processes have been presented in \cite{Kalinowski:2020rmb}, including VBF-type topologies. Cross sections were calculated using {\texttt{Madgraph5}}. 
We list the considered collider types and nominal center-of-mass energies as well as integrated luminosities in table \ref{tab:colls}.
\begin{center}
\begin{table}
\begin{center}
\begin{tabular}{c|c|c|c}
collider&cm energy [\TeV]&$\int\mathcal{L}$&{$\sigma_{_{1000}}$} [\fb]\\ \hline
%LHC now&13&$140\,\fb^{-1}$&7\\
HL-LHC&13/ 14&$3\,\ab^{-1}$&0.33\\
HE-LHC&27&$15\,\ab^{-1}$&0.07\\
FCC-hh&100&$20\,\ab^{-1}$&0.05\\ \hline
ee&3&$5\,\ab^{-1}$&0.2\\
$\mu\mu$&10&$10\,\ab^{-1}$&0.1\\
$\mu\mu$&30&$90\,\ab^{-1}$&0.01
\end{tabular}
\end{center}
\caption{Collider parameters used in the discovery reach {study} performed in \cite{Kalinowski:2020rmb}. Collider specifications have been taken from \cite{ATL-PHYS-PUB-2019-005,Collaboration:2650976,Abada:2019ono,Benedikt:2018csr,Delahaye:2019omf} for the HL-LHC, HE-LHC, FCC-hh and muon collider, respectively. The last column denotes the minimal cross section {required to produce} 1000 events using full target luminosity.}
\label{tab:colls}
\end{table}
\end{center}
We here label a scenario "realistic" when we can expect 1000 events to be produced using target luminosity and center-of-mass energies as specified above. Obviously, more detailed studies, including {both} background {contribution and detector response} simulation, are necessary to assess the actual collider reach.

\begin{figure}[htb]
\begin{center}%{%
\includegraphics[width=0.48\textwidth]{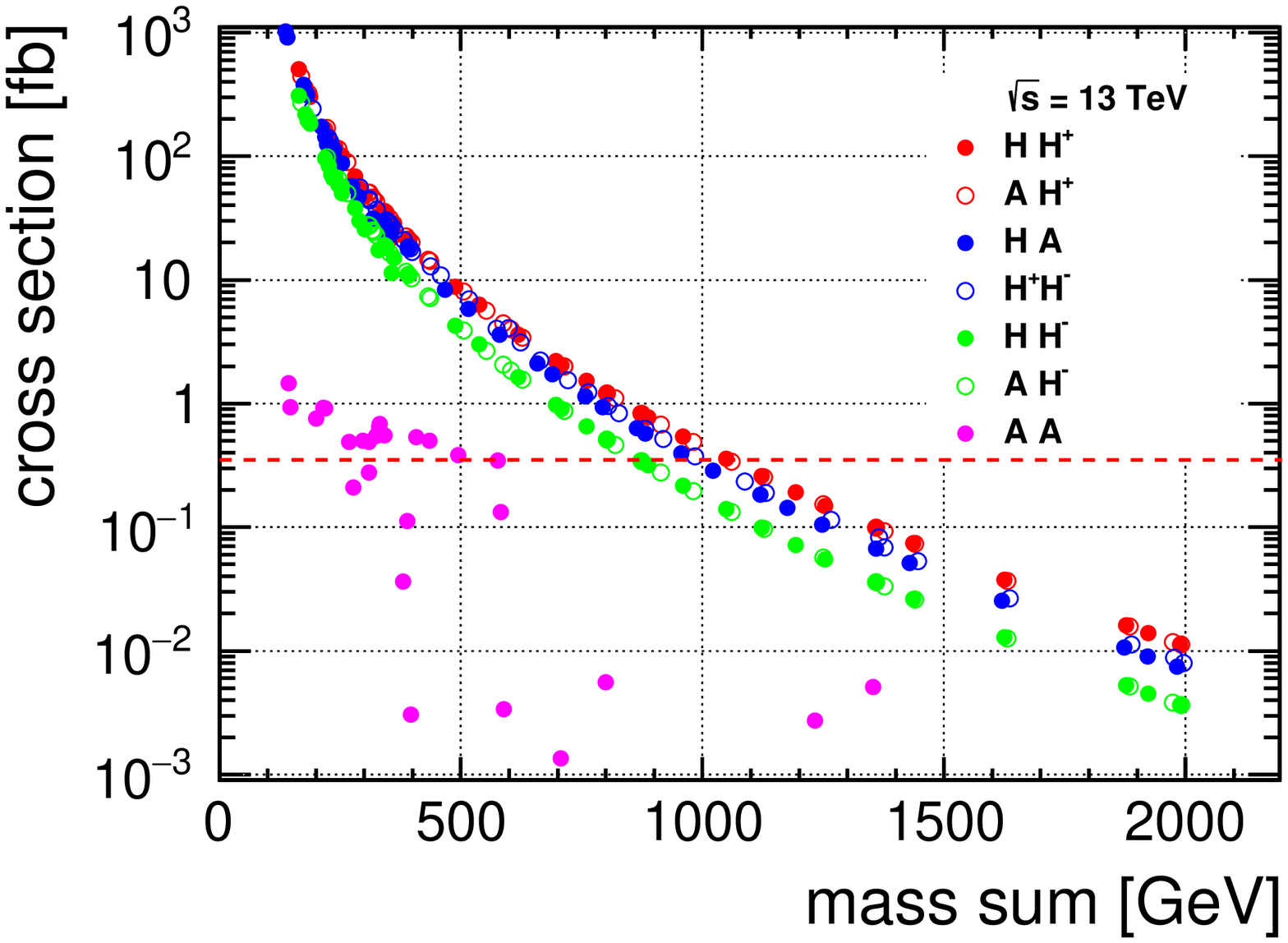}
\includegraphics[width=0.48\textwidth]{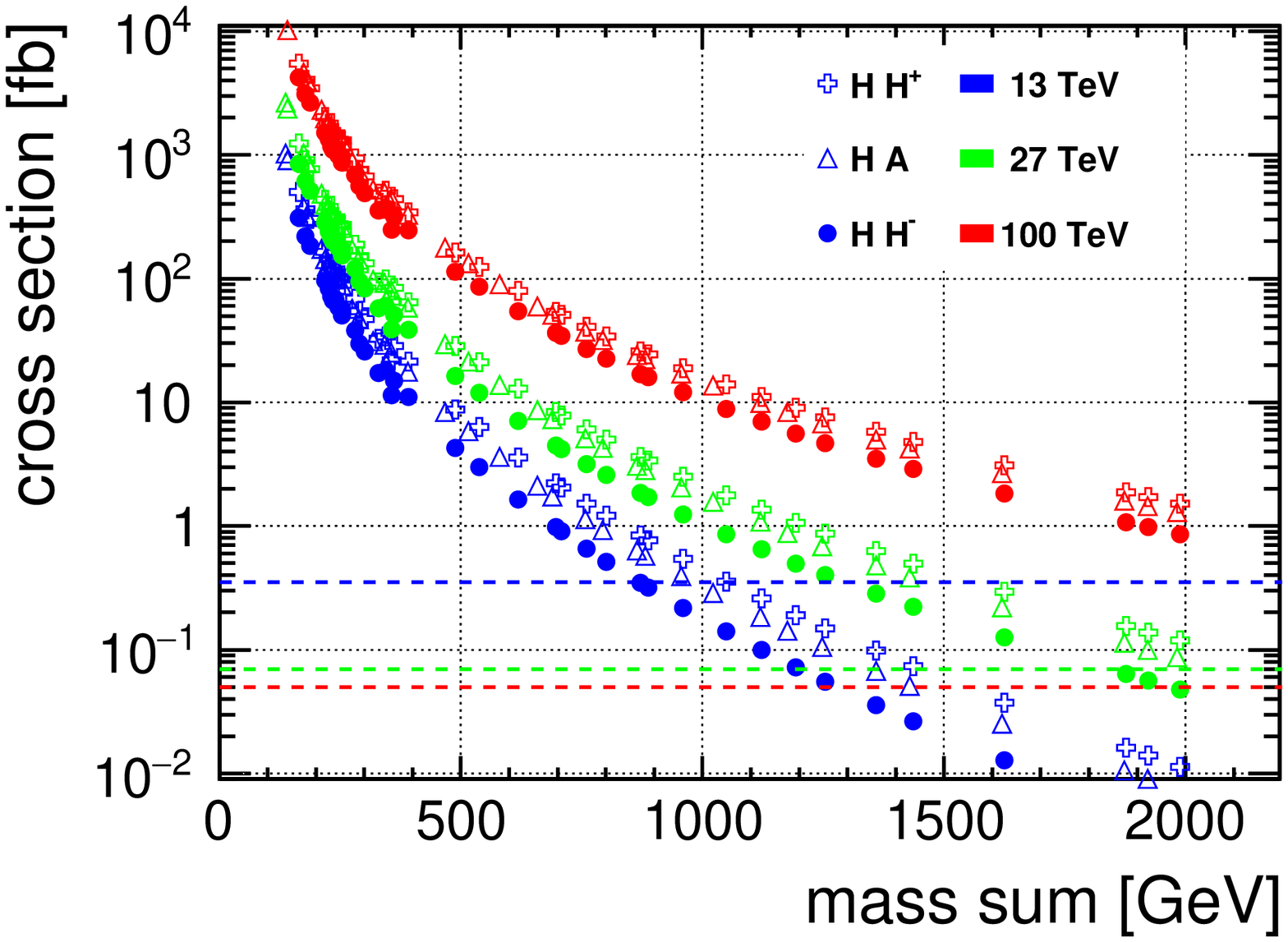}
\end{center}
%}
\caption{Pair-production cross-section predictions at $pp$ colliders as a function of the sum of produced particle {masses}. {\sl Left:} For {all considered production channels at} 13 \TeV LHC. {\sl Right:} {for selected channels} at 13\,\TeV, 27 \TeV, and 100 \TeV. {Horizontal dashed lines} denote the limit of the cross section at which 1000 events are produced with the respective target luminosity, cf table \ref{tab:colls}. Taken from \cite{Kalinowski:2020rmb}.}
\label{fig:pppair}
\end{figure}

\begin{figure}[htb]
\begin{center}%{%
\includegraphics[width=0.48\textwidth]{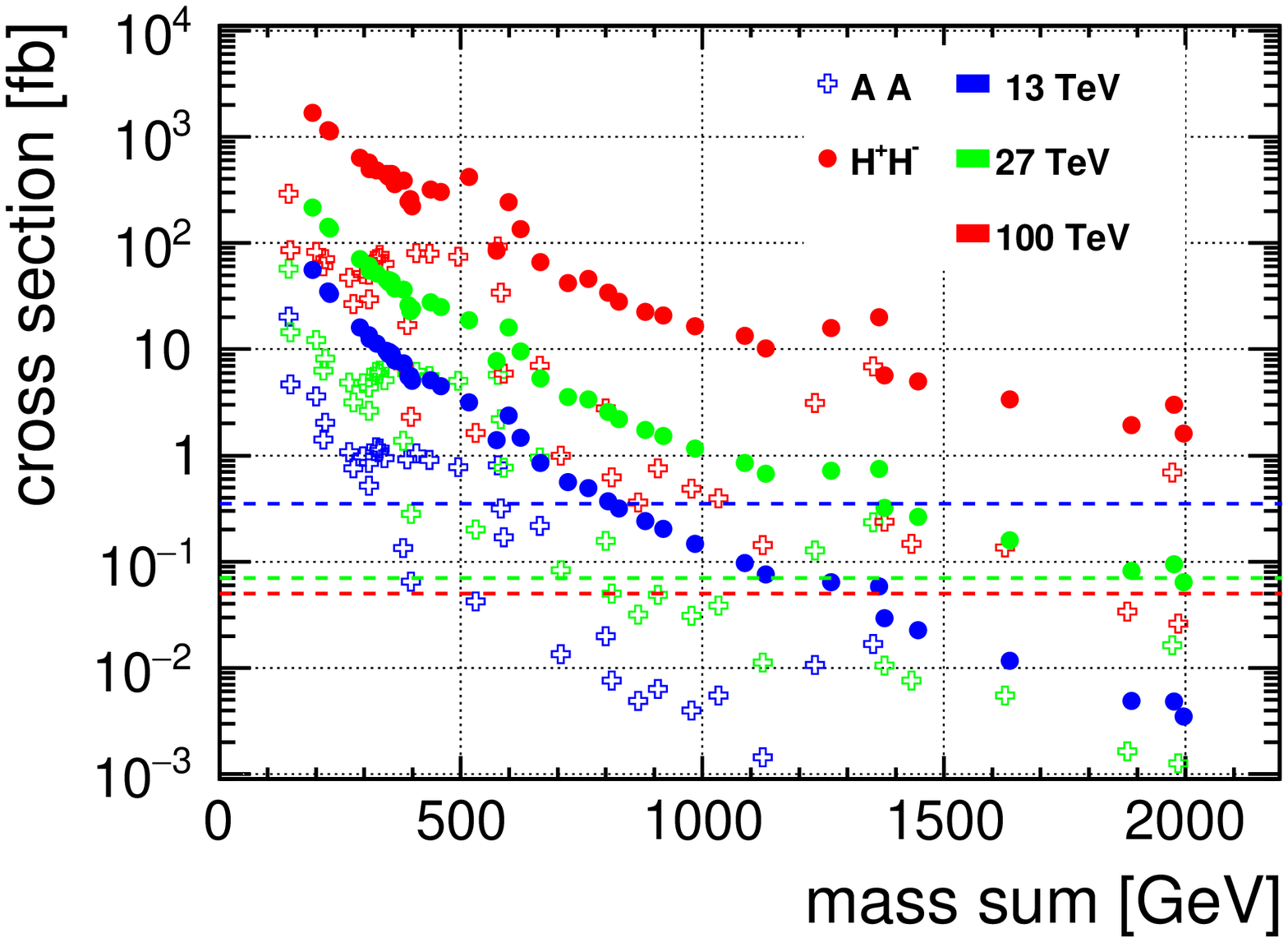}
\includegraphics[width=0.48\textwidth]{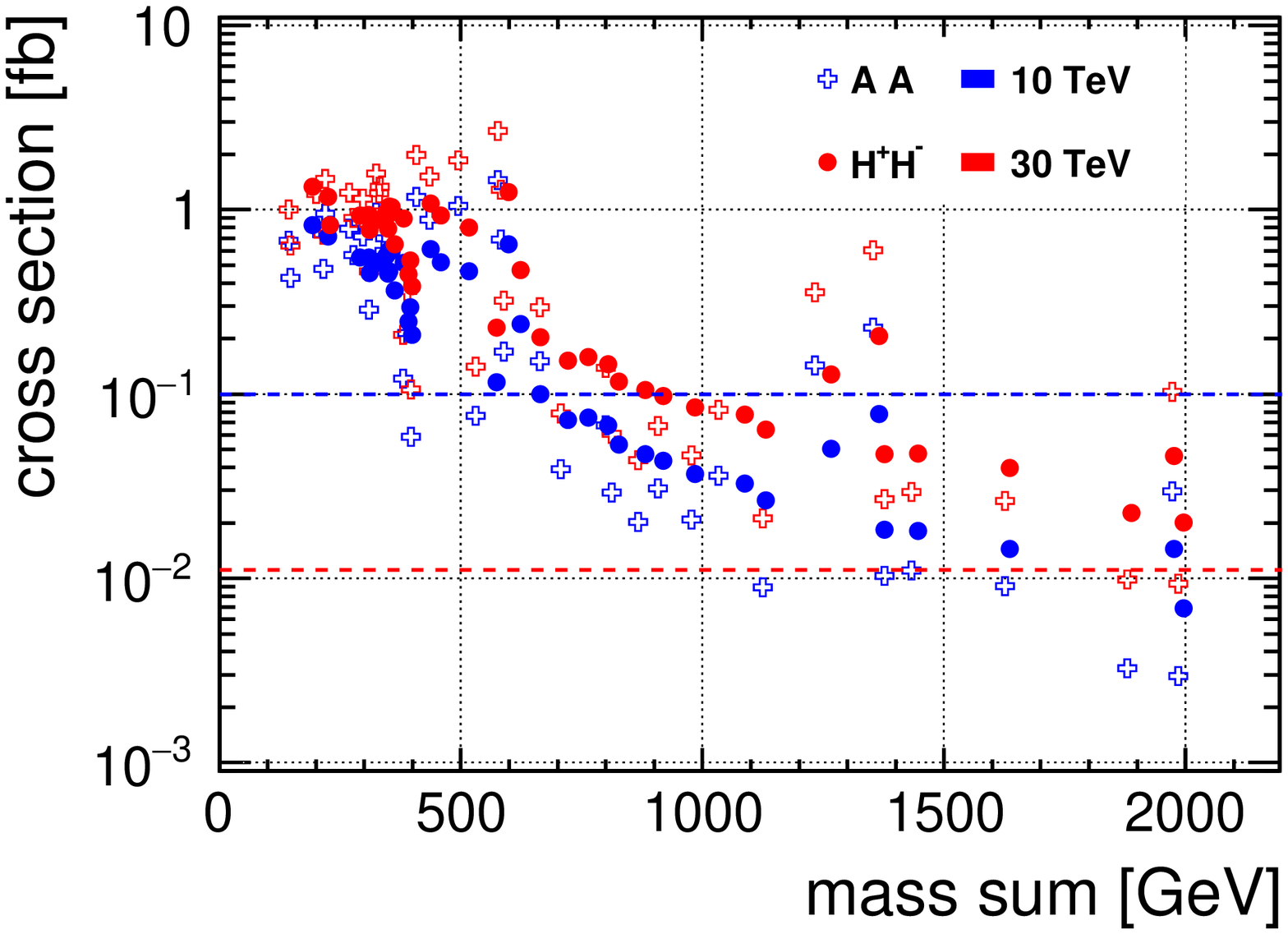}
\end{center}
%}
\caption{As figure \ref{fig:pppair}, but now considering the VBF-type production mode. {\sl Left:} for pp colliders, where two additional jets are produced and {\sl right:} at $\mu\mu$ colliders. Taken from \cite{Kalinowski:2020rmb}.}
\label{fig:vbf}
\end{figure}

We consider the following production modes:
\begin{itemize}
\item{}{\bf $pp$ colliders at various center-of-mass energies:}
\begin{eqnarray*}
p\,p&\rightarrow&H\,A,\,H\,H^+,\,H\,H^-,\,A\,H^+,\,A\,H^-,\,H^+\,H^-,\,AA,\\
p\,p&\rightarrow&A\,A\,j\,j,\,H^+\,H^-\,j\,j.
\end{eqnarray*}
The latter two processes are labelled "VBF-like" toplogies, although in practise we include all diagrams that contribute to that specific final state; e.g., to the $A\,A\,j\,j$ final state, also $H^\pm\,A$ production contributes, with subseqent decays $H^\pm\,\rightarrow\,W^\pm\,A$ and hadronic decays of the $W$. Furthermore, all but the $A\,A$ direct pair production are proportional to couplings from the SM electroweak sector (see e.g. \cite{Ilnicka:2015jba}), so in principle these production cross sections are determined by the masses of the pair-produced particles. The $AA$ channel is proportional to 
\begin{\eqn}\label{eqn:l345b}
\bar{\lam}_{345}\,=\,\lam_{345}-2\,\frac{M_H^2-M_A^2}{v^2},
\end{\eqn}
 so dependences here are more involved.
\item{\bf $\mu\mu$ colliders:}\\
At the muon collider, we mainly consider
\begin{\eqn*}
  \mu^+\,\mu^-\,\rightarrow\,\nu_\mu\,\bar{\nu}_\mu A A,\;\;\;
  \mu^+\,\mu^-\,\rightarrow\,\nu_\mu\,\bar{\nu}_\mu H^+ H^-.
\end{\eqn*}
which again corresponds to VBF-like production modes. However, as before, we do not specify intermediate states, so in fact {several} diagrams contribute which not all have a typical VBF topology. See appendix B and C of \cite{Kalinowski:2020rmb} for details.
\end{itemize}
Figures \ref{fig:pppair} and \ref{fig:vbf} show the production cross sections as a function of the mass sum of produced particles for various collider options and production modes. We see clearly that, while {predictions for} direct pair-production {cross sections} at $pp$ colliders exhibit a fall with rising mass-scales for all but the $AA$ pair-production mode, {understanding the behaviour of} the VBF-induced channels {is} less trivial. This can be attributed to the fact that more diagrams contribute. For example, for $AAjj$ we have contributions from $h-$ exchange in the s-channel which are again proportional to $\bar{\lam}_{345}$ given by eqn. (\ref{eqn:l345b}), which can induce large jumps between cross-section predictions for scenarios with similar mass scales. Similar differences can be observed for VBF-type production at $\mu\mu$ colliders; this can be traced back mainly to a fine-tuned cancellation of various contributing diagrams, which is discussed in large detail in \cite{Kalinowski:2020rmb}.

The summary of sensitivities in terms of mass scales is given in table \ref{tab:sens}.
\begin{center}
\begin{table}
\begin{center}
\begin{tabular}{||c||c||c||c||} \hline \hline
{collider}&{all others}& { $AA$} & {$AA$ +VBF}\\ \hline \hline
HL-LHC&1 \TeV&200-600 \GeV& 500-600 \GeV\\
HE-LHC&2 \TeV&400-1400 \GeV&800-1400 \GeV\\
FCC-hh&2 \TeV&600-2000 \GeV&1600-2000 \GeV\\ \hline \hline
CLIC, 3 \TeV&2 \TeV &- &300-600 \GeV\\
$\mu\mu$, 10 \TeV&2 \TeV &-&400-1400 \GeV\\
$\mu\mu$, 30 \TeV&2 \TeV  &-&1800-2000 \GeV \\ \hline \hline
\end{tabular}
\end{center}
\caption{Sensitivity of different collider options specified in table \ref{tab:colls}, using the sensitivity criterium of 1000 generated events in the specific channel. $x-y$ denotes minimal/ maximal mass scales that are reachable. Numbers for CLIC correspond to results from detailed investigations \cite{Kalinowski:2018kdn,deBlas:2018mhx}.}
\label{tab:sens}
\end{table}
\end{center}
We see that especially for $AA$ production the VBF mode at both proton and muon  colliders serves to significantly increase the discovery reach of the respective machine. Using the simple counting criterium above, we can furthermore state that a 27 \TeV proton-proton machine has a similar reach as a 10 \TeV muon collider, while 100 \TeV FCC-hh would correspond to a 30 \TeV muon-muon machine. Obviously, detailed investigations including SM background are needed to give a more realistic estimate of the respective collider reach.

\newpage
\section{Two-real-singlet extension}\label{sec:2rs}
\subsection{The model}
The two-real-singlet model (TRSM) is a model that features the extension of the SM scalar sector by two additional real singlets $S,\,X$, which obey a $\mathbb{Z}_2\,\otimes\,\mathbb{Z}_2'$ symmetry. It has been introduced in \cite{Robens:2019kga}, with a first phenomenological study of the model being presented in \cite{Papaefstathiou:2020lyp}. The model is also available in the public tool \texttt{ScannerS}~\cite{Coimbra:2013qq,Costa:2015llh,Muhlleitner:2020wwk}.

The transformation properties under the two discrete symmetries are specified as
\begin{equation}
    \begin{aligned}
        \Ztwo^S:\quad & S\to -S\eqcomma\ X\to X\eqcomma\ \text{SM} \to \text{SM}\eqcomma \\
        \Ztwo^X:\quad & X\to -X\eqcomma\ S \to S\eqcomma\ \text{SM} \to \text{SM}.
    \end{aligned}\label{eq:Z2syms}
\end{equation}
Application of this symmetry reduces the number of possible terms in the potential, such that we obtain 
\begin{equation}
    \begin{aligned}
        V & = \mu_{\Phi}^2 \Phi^\dagger \Phi + \lambda_{\Phi} {(\Phi^\dagger\Phi)}^2
        + \mu_{S}^2 S^2 + \lambda_S S^4
        + \mu_{X}^2 X^2 + \lambda_X X^4                                              \\
          & \quad+ \lambda_{\Phi S} \Phi^\dagger \Phi S^2
        + \lambda_{\Phi X} \Phi^\dagger \Phi X^2
        + \lambda_{SX} S^2 X^2\eqdot
    \end{aligned}\label{eq:TRSMpot}
\end{equation}
So far, we have not specified whether the additional scalar states acquire vevs. In fact, setting one of these to zero opens up the possibility of having a portal-like dark matter scenario. On the other hand, when both additional fields acquire a vev, the above symmetry is softly broken and all scalar fields mix. We here discuss this second scenario. The gauge-eigentstates are then given by
\begin{equation}
    \Phi = \begin{pmatrix} 0\\\frac{\phi_h + v}{\sqrt{2}}\end{pmatrix}\eqcomma\quad
    S = \frac{\phi_S + v_S}{\sqrt{2}}\eqcomma \quad
    X = \frac{\phi_X + v_X}{\sqrt{2}}
    \label{eq:fields}
\end{equation}
Rotation into mass eigenstates is then described by a rotation matrix $R$, with
\begin{equation}
    \begin{pmatrix}
        h_1 \\h_2\\h_3
    \end{pmatrix} = R \begin{pmatrix}
        \phi_h \\\phi_S\\\phi_X
    \end{pmatrix}\eqdot
\end{equation}
where in the following we adapt the convention that 
\begin{\eqn*}
M_1\,\leq\,M_2\,\leq\,M_3.
\end{\eqn*}
The rotation matrix is described via three mixing angles $\theta_{1,2,3}$, with 
\begin{equation}
    R = \begin{pmatrix}
        c_1 c_2             & -s_1 c_2             & -s_2     \\
        s_1 c_3-c_1 s_2 s_3 & c_1 c_3+ s_1 s_2 s_3 & -c_2 s_3 \\
        c_1 s_2 c_3+s_1 s_3 & c_1 s_3-s_1 s_2 c_3  & c_2 c_3
    \end{pmatrix}\eqdot
\end{equation}
with the short-hand notation
\begin{equation}
    s_1\equiv\sin\theta_{hS}\eqcomma\quad s_2\equiv\sin\theta_{hX}\eqcomma\quad s_3\equiv\sin\theta_{SX}\eqcomma\quad c_1\equiv\cos\theta_{hS}\eqcomma\ \ldots
\end{equation}
It is important to note that all interactions to SM particles are inherited through this mixing, with a corresponding scaling factor $\kappa_i\,\equiv\,R_{i1}$ for the mass eigenstate $h_i$.

After electroweak symmetry breaking, the model has in total 9 free parameters; as before, two of these, $v\,\simeq\,246\,\GeV$ and $M_a\,\simeq\,125\,\GeV$, are fixed by the Higgs mass measurement and electroweak precision observables. We then choose as free input parameters
\begin{equation}
    M_b\eqcomma\ M_c\eqcomma\ \theta_{hS}\eqcomma\ \theta_{hX}\eqcomma\ \theta_{SX}\eqcomma\ v_S\eqcomma\ v_X\eqcomma\label{eq:TRSMpars}
\end{equation}
{with $a\neq{}b\neq{}c\in\lbrace1,2,3\rbrace$.}

As before, the model is subject to a large number of theoretical and experimental constraints, which have been presented in detail in \cite{Robens:2019kga} and will not be repeated here.

\subsection{Phenomenology and benchmark planes}
Having three distinct scalar final states, this model allows for interesting scalar-scalar production and decay modes. At $pp$ colliders, we have
\begin{eqnarray*}
pp\,\rightarrow\,h_3\,\rightarrow\,h_1\,h_1;&& 
%\item{}
pp\,\rightarrow\,h_3\,\rightarrow\,h_2\,h_2;\\
%\item{}
pp\,\rightarrow\,h_2\,\rightarrow\,h_1\,h_1;&&
%\item{}
pp\,\rightarrow\,h_3\,\rightarrow\,h_1\,h_2
%\vspace{4mm}
\end{eqnarray*}
with decay modes given by
\begin{center}
$h_2\,\rightarrow\,\text{SM}$; 
%\item{}
$h_2\,\rightarrow\,h_1\,h_1$; $h_1\,\rightarrow\,\text{SM}$
\end{center}
The exact phenomenology depends on the chosen parameter point. While all partial decay widths to SM-like final states are given by the common scaling factors defined above 
\begin{equation}
    \Gamma(h_a\to\text{SM}; M_a) = \kappa_a^2 \cdot \Gamma_\text{tot}(h_\text{SM}; M_a),\label{eq:widthscaling}
\end{equation}
{where $ \Gamma_\text{tot}(h_\text{SM}; M_a)$ denotes the decay width of a SM-like scalar of mass $M_a$,}
partial decays into scalar final states need to be calculated from the new physics parameters in the potential. 

In \cite{Robens:2019kga}, a number of benchmark planes was defined in order to accomodate for production and decay modes in the scalar sector that are currently not investigated by the LHC experiments. We list these in table \ref{tab:benchmarkoverview}.
\begin{center}
\begin{table}
    \centering
%    \begin{tabular}{\textwidth}{sssb}
{\small
\begin{tabular}{cccc}
        \hline \\
        benchmark scenario & $h_{125}$ & target signature      & possible successive decays                          \\
        \hline \\
        \textbf{BP1}       & $h_3$               & $h_{125} \to h_1 h_2$ & $h_2 \to h_1 h_1$ if $M_2 > 2 M_1$                  \\
        \textbf{BP2}       & $h_2$               & $h_3 \to h_1 h_{125}$ & -                                                   \\
        \textbf{BP3}       & $h_1$               & $h_3 \to h_{125} h_2$ & $h_2 \to h_{125}h_{125}$ if $M_2 > \SI{250}{\GeV}$  \\
        \textbf{BP4}       & $h_3$               & $h_2 \to h_1 h_1$     & -                                                   \\
        \textbf{BP5}       & $h_2$               & $h_3 \to h_1 h_1$     & -                                                   \\
        \textbf{BP6}       & $h_1$               & $h_3 \to h_2 h_2$     & $h_2 \to h_{125}h_{125}$ if  $M_2 > \SI{250}{\GeV}$ \\
        \hline
    \end{tabular}
}
    \caption{Overview of the benchmark scenarios: The second column denotes the
    Higgs mass eigenstate that we identify with the observed Higgs boson,
    $h_{125}$, the third column names the targeted decay mode of the resonantly
    produced Higgs state, and the fourth column lists possible relevant
    successive decays of the resulting Higgs states. Taken from \cite{Robens:2019kga}.}\label{tab:benchmarkoverview}
\end{table}
\end{center}

For some of these, especially BP4, BP5, and BP1, relatively high production rates of 60 \pb, 2.5 \pb, and 3 \pb respectively can be achieved at the 13 \TeV LHC. These numbers correspond to rescaled production modes
 \begin{equation}\label{eqn:sigfac}
    \sigma(M_a) = \kappa_a^2 \cdot \sigma_\text{SM} ( M_a).
\end{equation}
followed by factorized decays. $\sigma_\text{SM} ( M_a)$ denotes the NNLO+NNLL production cross section for a SM-like Higgs of mass $M_a$, with numbers taken from \cite{Heinemeyer:2013tqa}. The relatively large rates make these BPs prime targets for current LHC data analyses. In the following, we however concentrate on BP3 and the $hhh$ final state.

\subsection{$hhh$ production in the TRSM}
\begin{figure}[htb]
\begin{center}%{%
\includegraphics[width=0.48\textwidth]{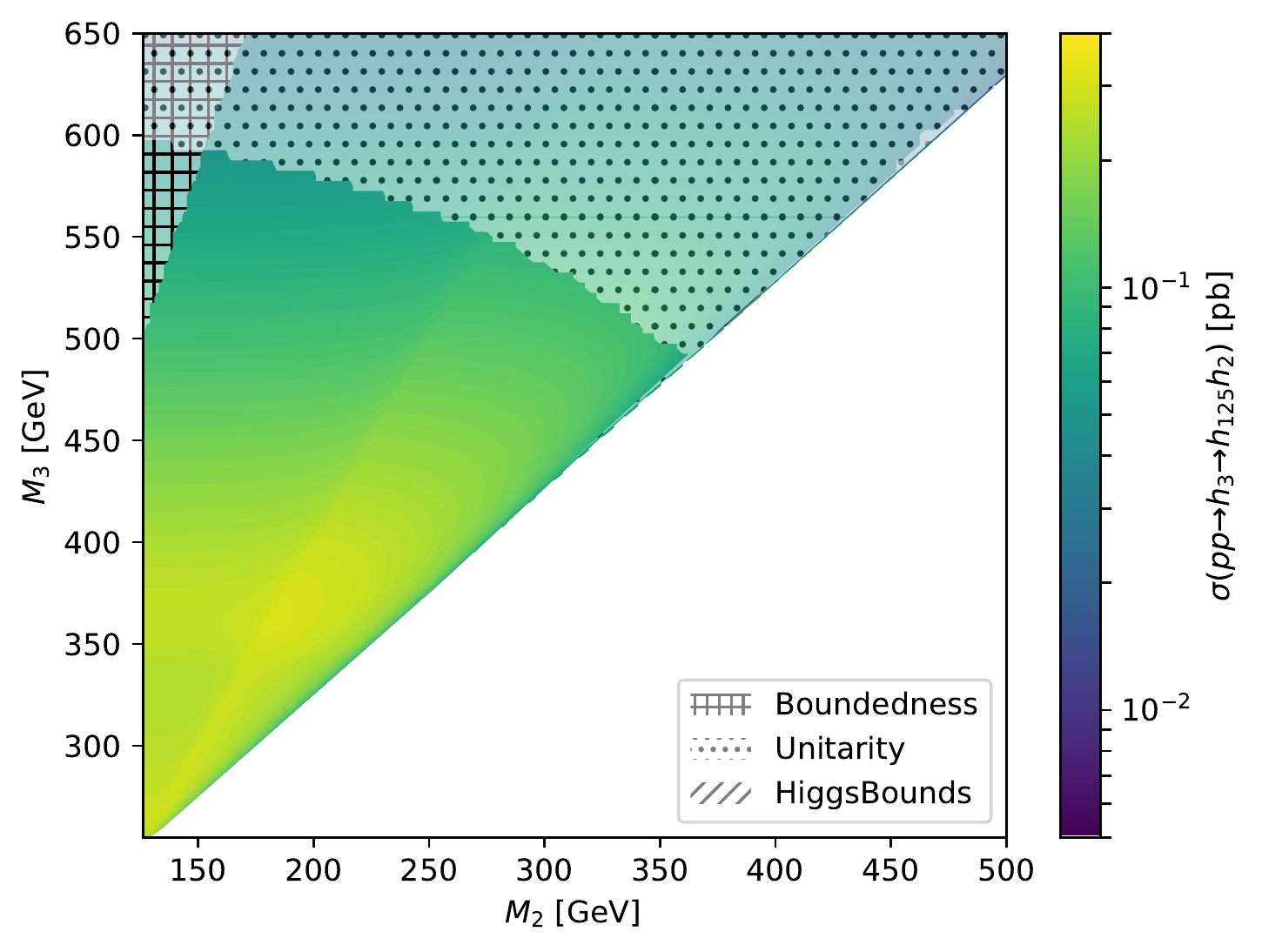}
\end{center}
%}
\caption{Production cross section for the ${h}h_2$ final state in BP3 at a 13 \TeV LHC, as a function of $M_2$ and $M_3$. Experimental exclusion bounds stem from searches for $h_{2,3}\,\rightarrow\,V\,V$ from 2016 LHC Run II data \cite{Aaboud:2017rel,Sirunyan:2018qlb,Aaboud:2018bun}. Taken from \cite{Robens:2019kga}.}
\label{fig:bp3}
\end{figure}

\begin{figure}[htb]
\begin{center}%{%
\begin{minipage}{0.44\textwidth}
\includegraphics[width=\textwidth]{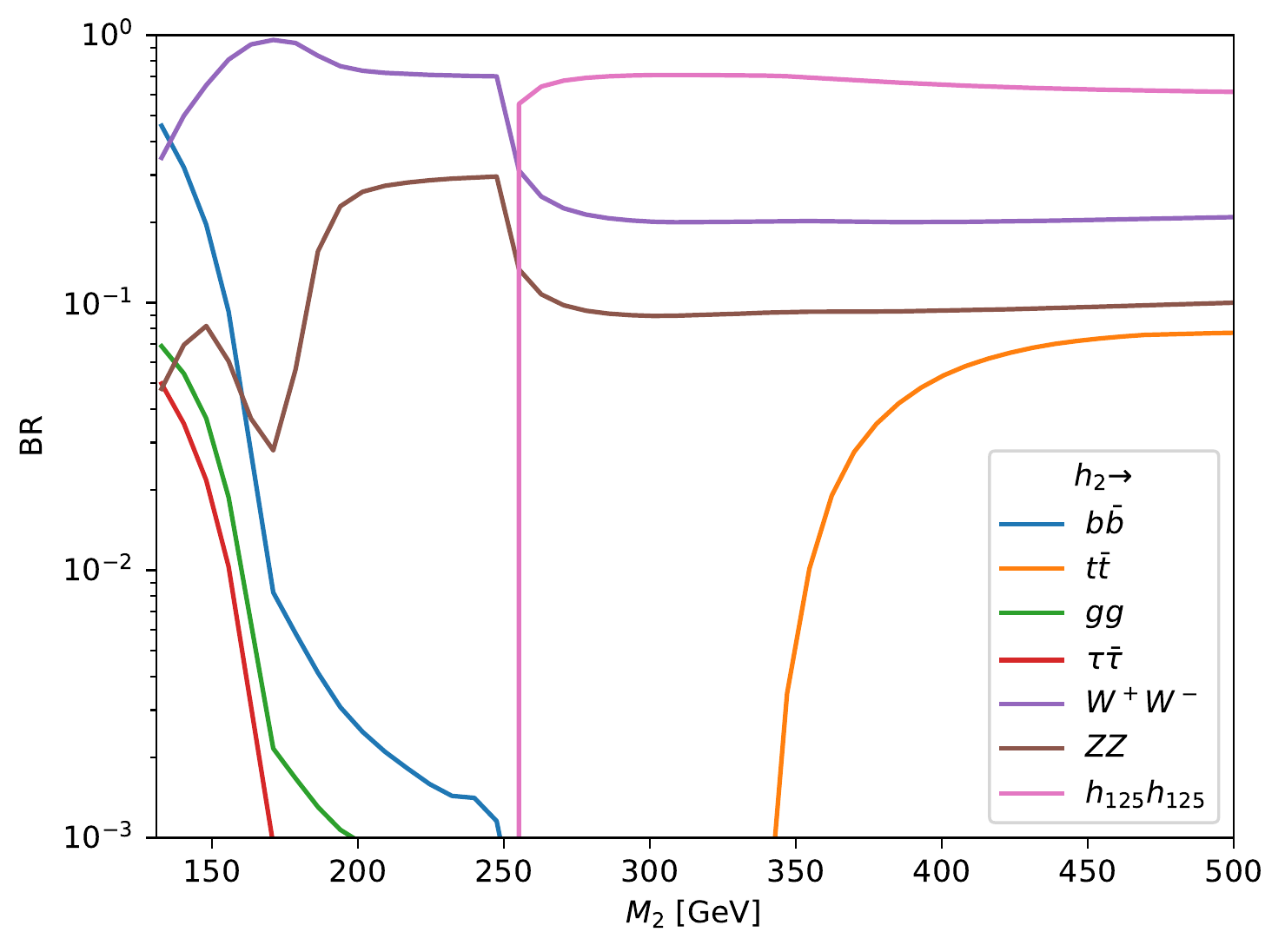}
\end{minipage}
\begin{minipage}{0.52\textwidth}
\includegraphics[width=\textwidth]{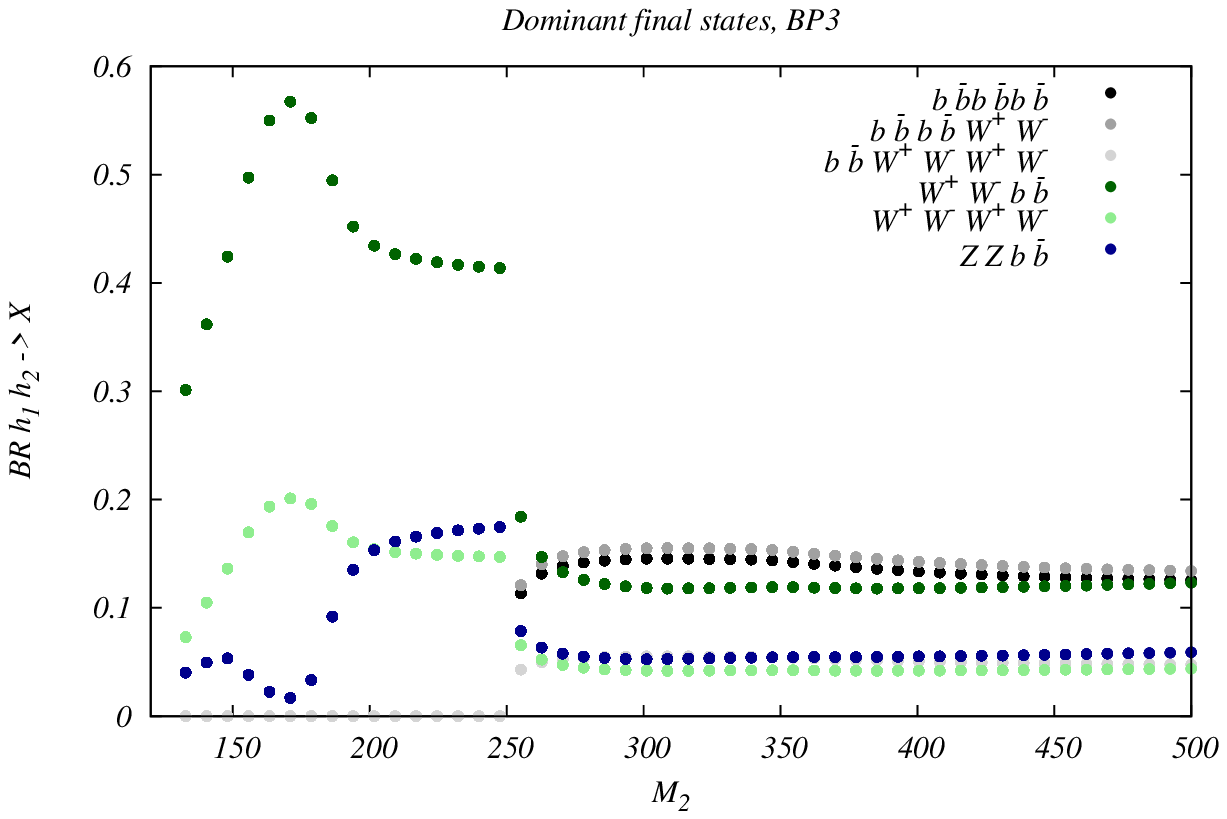}
\end{minipage}
\end{center}
%}
\caption{ Branching ratios for $h_2$ {\sl (left)} and ${h}\,h_2$ {\sl (right)} in BP3, as a function of $M_2$. Left plot taken from \cite{Robens:2019kga}.}
\label{fig:BRs}
\end{figure}

One interesting scenario within the TRSM is the asymmetric production and subsequent decay
\begin{equation}\label{eq:chain}
p p \,\rightarrow\, h_3\,\rightarrow\,h_2\,h_1\,\rightarrow\,h_1\,h_1
\,h_1,
\end{equation}
where $h_1\,\equiv\,{h}$ is the SM-like scalar. This signature is realized in BP3 and was analysed in detail in \cite{Papaefstathiou:2020lyp}. Input parameters for BP3 are displayed in table \ref{tab:BP3}.
\begin{center}
\begin{table}
\centering
\begin{tabular}{cc}
Parameter & Value \\	
\hline
$M_1$ & $125.09~\rm{GeV}$\\
$M_2$ & $[125,~500]~\rm{GeV}$\\
$M_3$ & $[255,~650]~\rm{GeV}$\\
$\theta_{hS}$ & $-0.129$\\
$\theta_{hX}$ & $0.226$ \\
$\theta_{SX}$ & $-0.899$\\
$v_S$ & $140~\rm{GeV}$\\
$v_X$ & $100~\rm{GeV}$\\ \hline
$\kappa_1$ & $0.966$\\
$\kappa_2$ & $0.094$\\
$\kappa_3$ & $0.239$
\end{tabular}	
\caption{\label{tab:BP3} The numerical values for the independent parameter values of eq.~(\ref{eq:TRSMpars}) that characterise \textbf{BP3}. The Higgs doublet vev, $v$, is fixed to 246~GeV. The $\kappa_i$ values correspond to the rescaling parameters of the SM-like couplings for the respective scalars and are derived quantities.}
\end{table}
\end{center}
We want to briefly comment on the calculation of rates, and differences between these in \cite{Robens:2019kga} and \cite{Papaefstathiou:2020lyp}:
\begin{itemize}
\item{}In \cite{Robens:2019kga}, production cross sections were calculated as specified in eqn. (\ref{eqn:sigfac}), and rates for final states were then derived via multiplication with the corresponding branching ratios. This is a priori a good approach for a first estimate, and incorporates important higher-order effects in the production cross sections.
\item{}In \cite{Papaefstathiou:2020lyp}, on the other hand, we made use of a customized  \texttt{loop\_sm} model implemented in \texttt{MadGraph5\_aMC@NLO} (v2.7.3)~\cite{Alwall:2014hca, Hirschi:2015iia}, and subsequently interfaced to \texttt{HERWIG} (v7.2.1)~\cite{Bahr:2008pv, Gieseke:2011na, Arnold:2012fq, Bellm:2013hwb, Bellm:2015jjp, Bellm:2017bvx, Bellm:2019zci}. This model includes full top and bottom mass effects and calculates production modes at LO, i.e. at the one-loop level. Furthermore, for the process (\ref{eq:chain}) intermediate states were not specified, which guarantees the inclusion of all contributing diagrams as well as interference effects. In particular, in some scenarions contributions from $s$-channel offshell $h_1$ states were on the $\%$ level.
\end{itemize}
Taking this into account, we display in figure \ref{fig:bp3} the production cross section for the ${h}\,h_2$ final state as derived in \cite{Robens:2019kga}. For this final state, production cross sections can reach up to 0.3 \pb.
Branching ratios for $h_2$ and ${h}\,h_2$ are displayed in figure \ref{fig:BRs}.
This particular benchmark plane was chosen such that $h_2\,\rightarrow\,{h}{h}$ becomes dominant as soon as it is kinematically allowed. This induces the predominance of $b\bar{b}b\bar{b}b\bar{b}$ and $b\bar{b}b\bar{b}W^+W^-$ rates over those for four-particle final states.

In \cite{Papaefstathiou:2020lyp}, several benchmark points were selected which were then investigated at the LHC using 14 \TeV center-of-mass energy, where we concentrated in the $b\bar{b}b\bar{b}b\bar{b}$ final state. Those benchmark points as well as significances for an integrated luminosity of $\int\,\mathcal{L}\,=\,300\,\fb^{-1}$ and  $\int\,\mathcal{L}\,=\,3000\,\fb^{-1}$ are shown in table \ref{tab:hhh}.
\begin{center}
\begin{table}
{\small
\begin{center}
\begin{tabular}{c||cc||cc}\\
{\bf $(M_2, M_3)$}& $\sigma(pp\rightarrow h_1 h_1 h_1)$ &
$\sigma(pp\rightarrow 3 b \bar{b})$&$\text{sig}|_{300\rm{fb}^{-1}}$& $\text{sig}|_{3000\rm{fb}^{-1}}$\\
${[\GeV]}$ & ${[\fb]}$  & ${[\fb]}$ & &\\
\hline\hline
$(255, 504)$ & $32.40$ & $6.40$&$2.92$&{  $9.23$}\\
$(263, 455)$ & $50.36$ & $9.95$&{ $4.78$}&{  $15.10 $}\\
$(287, 502)$ & $39.61$ & $7.82$&{  $4.01$} &{  $12.68$}\\
$(290, 454)$ & $49.00$ & $9.68$&{  $5.02$}&{  $15.86 $}\\
$(320, 503)$ & $35.88$& $7.09$& {  $3.76 $}&{  $11.88$}\\
$(264, 504)$ & $37.67$ & $7.44$&{  $3.56 $}&{  $11.27 $}\\
$(280, 455)$& $51.00$ & $10.07$&{  $5.18$} &{  $16.39$}\\
$(300, 475)$&$43.92$& $8.68$&{  $4.64 $}&{  $14.68 $}\\
$(310, 500)$& $37.90$ & $7.49$&{  $4.09 $}&{  $12.94$}\\
$(280, 500)$& $40.26$& $7.95$&{  $4.00 $}&{  $12.65 $}\\
\end{tabular}
\end{center}}
\caption{Benchmark points investigated in \cite{Papaefstathiou:2020lyp}, {leading-order} production cross sections at 14 \TeV, as well as significances for different integrated luminosities.}
\label{tab:hhh}
\end{table}
\end{center}
For details of the analysis as well as SM background simulation, we refer the reader to the above work. We see that several of the benchmark points are in the 4-5 $\sigma$ range already for a relatively low luminosity, and all have significances above the discovery reach after the full run of HL-LHC. We therefore strongly encourage the experimental collaborations to adapt our search stragety, using actual LHC data.

\begin{center}
\begin{figure}[htb]
\begin{center}%{%
\includegraphics[width=0.48\textwidth]{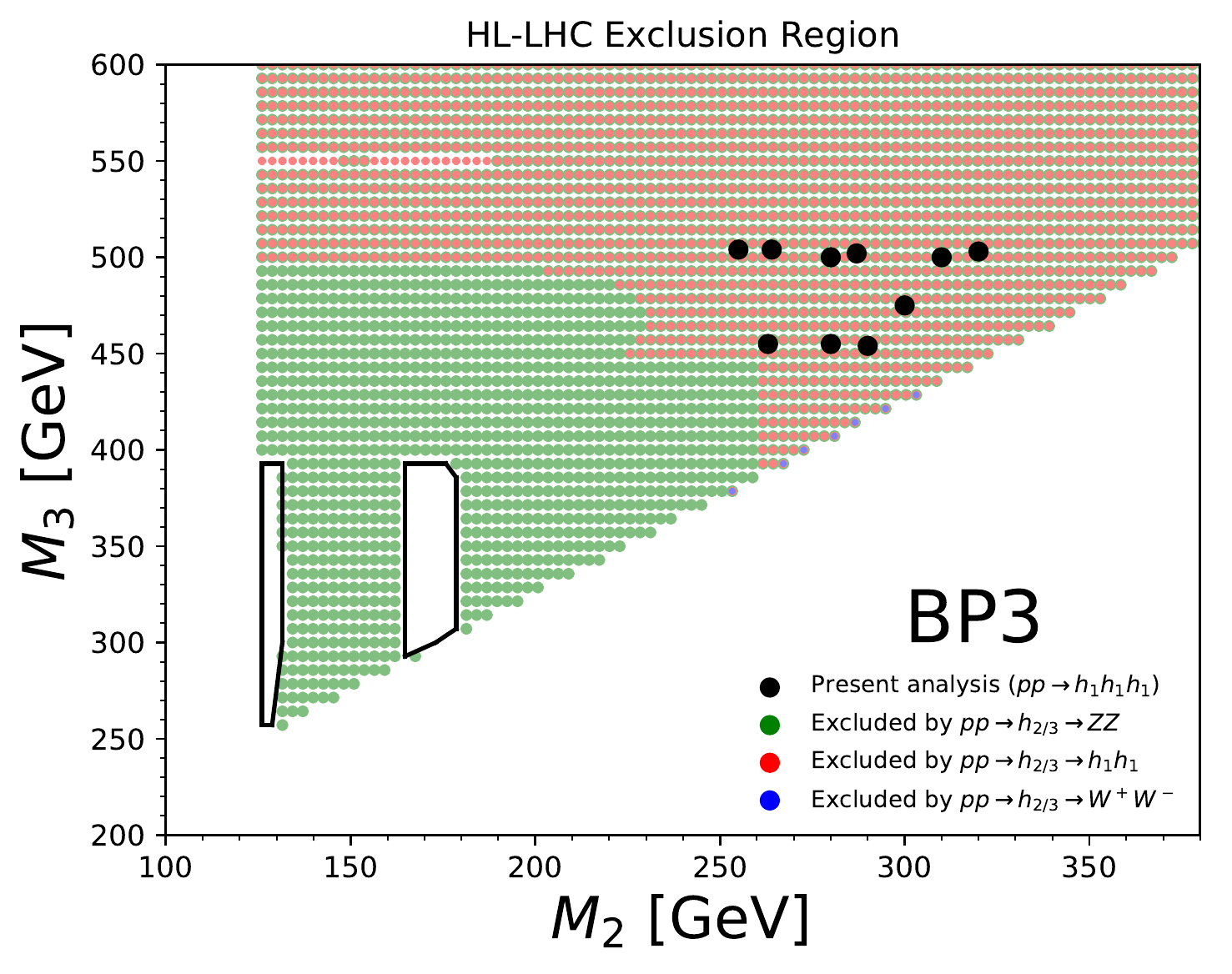}
\end{center}
\caption{Constraints on the $\lb M_2,\,M_3 \rb$ plane from extrapolation of other searches at the HL-LHC from extrapolation (see text for details). Taken from \cite{Papaefstathiou:2020lyp}.}
\label{fig:hlothers}
\end{figure}
\end{center}

Finally, we can ask whether other channels can not equally constrain the allowed parameter space at the HL-LHC. To this end, we have extrapolated various analyses assessing the heavy Higgs boson prospects of the HL-LHC in final states originating from $h_i \rightarrow h_1 h_1$ \cite{Sirunyan:2018two,Aad:2019uzh}, $h_i \rightarrow ZZ$ \cite{Sirunyan:2018qlb,Cepeda:2019klc} and $h_i \rightarrow W^+W^-$ \cite{Aaboud:2017gsl,ATL-PHYS-PUB-2018-022}, for $i=2,3$, and combined these with extrapolations of results from 13 TeV where appropriate. Details of the extrapolation procedure can be found in Appendix~D of ref.~\cite{Papaefstathiou:2020iag}. The corresponding results are shown in figure \ref{fig:hlothers}.

Especially $ZZ$ final states can probe nearly all of the available parameter space. However, we want to emphasize that these depend on different model parameters than the $h_1\,h_1\,h_1$ final state rates, and therefore these searches can be considered as complementary, testing various parts of the new physics potential. We encourage the LHC experimental collaborations to pursue searches in all possible decay channels.

\section{Conclusion}\label{sec:concl}

In this work, we have discussed several new physics models and their discovery prospects at future colliders. For the Inert Doublet Model, a two-Higgs doublet model with a dark matter candidate, we have discussed constraints on the parameter space and presented results of a dedicated study of several benchmark points in di-scalar production within the CLIC environment for several center-of-mass energies. We found that, when using both leptonic and semi-leptonic decay modes for these searches, mass scales up to 1 \TeV could in principle be reachable. Furthermore, we have performed a simplified study where we compared collider reaches for IDM benchmark points using a simple counting criterium. We found that, especially for the elusive $AA$ production channel, the inclusion of VBF-type production modes greatly enhances the mass range that could in principle be tested at colliders. This comparison should be taken as a starting point for more dedicated studies at both hadron-hadron and lepton-lepton machines.

Furthermore, we reported on the sensitivity of the HL-LHC for {new physics-}initiated cascade decays leading to ${h}\,{h}\,{h}$ final states, with subsequent decays into 3 $b\,\bar{b}$ pairs. We found that at the HL-LHC all considered benchmark points such be within discovery range, while some statements could already be made for several benchmark points for an integrated luminosity of a few hundred $\fb^{-1}$. We strongly encourage the LHC experimental collaborations to adapt this search using current and future LHC data.

\section*{Acknowledgements}
This research was supported in parts by the National Science Centre, Poland, the
HARMONIA project under contract UMO-2015/18/M/ST2/00518 and
OPUS project under contract UMO-2017/25/B/ST2/00496 (2018-2021), {and by the European Union through the Programme Horizon 2020 via the COST actions CA15108 - FUNDAMENTALCONNECTIONS and CA16201 - PARTICLEFACE.} 
Research discussed here was also supported by the UK's Royal Society.

%uncomment the following lines to place a figure
%\begin{figure}[htb]
%\centerline{%
%\includegraphics[width=12.5cm]{Fig1}}
%\caption{Plot of ...}
%\label{Fig:F2H}
%\end{figure}
%\bibliographystyle{hunsrt}
%\bibliography{lit,two_scalars}

\begin{thebibliography}{10}

\bibitem{Aad:2012tfa}
Georges Aad et~al.
\newblock {Observation of a new particle in the search for the Standard Model
  Higgs boson with the ATLAS detector at the LHC}.
\newblock {\em Phys. Lett.}, B716:1--29, 2012, 1207.7214.

\bibitem{Chatrchyan:2012ufa}
Serguei Chatrchyan et~al.
\newblock {Observation of a New Boson at a Mass of 125 GeV with the CMS
  Experiment at the LHC}.
\newblock {\em Phys. Lett.}, B716:30--61, 2012, 1207.7235.

\bibitem{atlpub}
https://twiki.cern.ch/twiki/bin/view/AtlasPublic/HiggsPublicResults.

\bibitem{cmspub}
http://cms-results.web.cern.ch/cms-results/public-results/publications/HIG/index.html.

\bibitem{Deshpande:1977rw}
Nilendra~G. Deshpande and Ernest Ma.
\newblock {Pattern of Symmetry Breaking with Two Higgs Doublets}.
\newblock {\em Phys. Rev.}, D18:2574, 1978.

\bibitem{Cao:2007rm}
Qing-Hong Cao, Ernest Ma, and G.~Rajasekaran.
\newblock {Observing the Dark Scalar Doublet and its Impact on the
  Standard-Model Higgs Boson at Colliders}.
\newblock {\em Phys. Rev.}, D76:095011, 2007, 0708.2939.

\bibitem{Barbieri:2006dq}
Riccardo Barbieri, Lawrence~J. Hall, and Vyacheslav~S. Rychkov.
\newblock {Improved naturalness with a heavy Higgs: An Alternative road to LHC
  physics}.
\newblock {\em Phys. Rev.}, D74:015007, 2006, hep-ph/0603188.

\bibitem{Ilnicka:2015jba}
Agnieszka Ilnicka, Maria Krawczyk, and Tania Robens.
\newblock {Inert Doublet Model in light of LHC Run I and astrophysical data}.
\newblock {\em Phys. Rev.}, D93(5):055026, 2016, 1508.01671.

\bibitem{Ilnicka:2018def}
Agnieszka Ilnicka, Tania Robens, and Tim Stefaniak.
\newblock {Constraining Extended Scalar Sectors at the LHC and beyond}.
\newblock {\em Mod. Phys. Lett.}, A33(10n11):1830007, 2018, 1803.03594.

\bibitem{Dercks:2018wch}
Daniel Dercks and Tania Robens.
\newblock {Constraining the Inert Doublet Model using Vector Boson Fusion}.
\newblock {\em Eur. Phys. J.}, C79(11):924, 2019, 1812.07913.

\bibitem{Kalinowski:2018ylg}
Jan Kalinowski, Wojciech Kotlarski, Tania Robens, Dorota Sokolowska, and
  Aleksander~Filip Zarnecki.
\newblock {Benchmarking the Inert Doublet Model for $e^+ e^-$ colliders}.
\newblock {\em JHEP}, 12:081, 2018, 1809.07712.

\bibitem{Kalinowski:2020rmb}
Jan Kalinowski, Tania Robens, Dorota Sokolowska, and Aleksander~Filip Zarnecki.
\newblock {IDM benchmarks for the LHC and future colliders}.
\newblock 2020, 2012.14818.

\bibitem{Eriksson:2009ws}
David Eriksson, Johan Rathsman, and Oscar St{\aa}l.
\newblock {2HDMC: Two-Higgs-Doublet Model Calculator Physics and Manual}.
\newblock {\em Comput. Phys. Commun.}, 181:189--205, 2010, 0902.0851.

\bibitem{Bechtle:2008jh}
Philip Bechtle, Oliver Brein, Sven Heinemeyer, Georg Weiglein, and Karina~E.
  Williams.
\newblock {HiggsBounds: Confronting Arbitrary Higgs Sectors with Exclusion
  Bounds from LEP and the Tevatron}.
\newblock {\em Comput. Phys. Commun.}, 181:138--167, 2010, 0811.4169.

\bibitem{Bechtle:2011sb}
Philip Bechtle, Oliver Brein, Sven Heinemeyer, Georg Weiglein, and Karina~E.
  Williams.
\newblock {HiggsBounds 2.0.0: Confronting Neutral and Charged Higgs Sector
  Predictions with Exclusion Bounds from LEP and the Tevatron}.
\newblock {\em Comput. Phys. Commun.}, 182:2605--2631, 2011, 1102.1898.

\bibitem{Bechtle:2013wla}
Philip Bechtle, Oliver Brein, Sven Heinemeyer, Oscar St{\aa}l, Tim Stefaniak,
  Georg Weiglein, and Karina~E. Williams.
\newblock {$\mathsf{HiggsBounds}-4$: Improved Tests of Extended Higgs Sectors
  against Exclusion Bounds from LEP, the Tevatron and the LHC}.
\newblock {\em Eur. Phys. J.}, C74(3):2693, 2014, 1311.0055.

\bibitem{Bechtle:2015pma}
Philip Bechtle, Sven Heinemeyer, Oscar St{\aa}l, Tim Stefaniak, and Georg
  Weiglein.
\newblock {Applying Exclusion Likelihoods from LHC Searches to Extended Higgs
  Sectors}.
\newblock {\em Eur. Phys. J.}, C75(9):421, 2015, 1507.06706.

\bibitem{Bechtle:2020pkv}
Philip Bechtle, Daniel Dercks, Sven Heinemeyer, Tobias Klingl, Tim Stefaniak,
  Georg Weiglein, and Jonas Wittbrodt.
\newblock {HiggsBounds-5: Testing Higgs Sectors in the LHC 13 TeV Era}.
\newblock {\em Eur. Phys. J.}, C80(12):1211, 2020, 2006.06007.

\bibitem{Bechtle:2013xfa}
Philip Bechtle, Sven Heinemeyer, Oscar Stal, Tim Stefaniak, and Georg Weiglein.
\newblock {$HiggsSignals$: Confronting arbitrary Higgs sectors with
  measurements at the Tevatron and the LHC}.
\newblock {\em Eur. Phys. J.}, C74(2):2711, 2014, 1305.1933.

\bibitem{Bechtle:2020uwn}
Philip Bechtle, Sven Heinemeyer, Tobias Klingl, Tim Stefaniak, Georg Weiglein,
  and Jonas Wittbrodt.
\newblock {HiggsSignals-2: Probing new physics with precision Higgs
  measurements in the LHC 13 TeV era}.
\newblock {\em Eur. Phys. J.}, C81(2):145, 2021, 2012.09197.

\bibitem{Belanger:2020gnr}
Genevieve Belanger, Ali Mjallal, and Alexander Pukhov.
\newblock {Recasting direct detection limits within micrOMEGAs and implication
  for non-standard Dark Matter scenarios}.
\newblock {\em Eur. Phys. J.}, C81(3):239, 2021, 2003.08621.

\bibitem{Alwall:2011uj}
Johan Alwall, Michel Herquet, Fabio Maltoni, Olivier Mattelaer, and Tim
  Stelzer.
\newblock {MadGraph 5 : Going Beyond}.
\newblock {\em JHEP}, 06:128, 2011, 1106.0522.

\bibitem{Goudelis:2013uca}
A.~Goudelis, B.~Herrmann, and O.~Stal.
\newblock {Dark matter in the Inert Doublet Model after the discovery of a
  Higgs-like boson at the LHC}.
\newblock {\em JHEP}, 09:106, 2013, 1303.3010.

\bibitem{ufo_idm}
https://feynrules.irmp.ucl.ac.be/wiki/ModelDatabaseMainPage.
\newblock (as checked on Dec. 18,2020).

\bibitem{Zyla:2020zbs}
P.~A. Zyla et~al.
\newblock {Review of Particle Physics}.
\newblock {\em PTEP}, 2020(8):083C01, 2020.

\bibitem{gfitter}
http://project-gfitter.web.cern.ch/project-gfitter/.

\bibitem{Haller:2018nnx}
Johannes Haller, Andreas Hoecker, Roman Kogler, Klaus Moenig, Thomas Peiffer,
  and Joerg Stelzer.
\newblock {Update of the global electroweak fit and constraints on
  two-Higgs-doublet models}.
\newblock {\em Eur. Phys. J.}, C78(8):675, 2018, 1803.01853.

\bibitem{Aghanim:2018eyx}
N.~Aghanim et~al.
\newblock {Planck 2018 results. VI. Cosmological parameters}.
\newblock {\em Astron. Astrophys.}, 641:A6, 2020, 1807.06209.

\bibitem{Aprile:2018dbl}
E.~Aprile et~al.
\newblock {Dark Matter Search Results from a One Tonne$\times$Year Exposure of
  XENON1T}.
\newblock {\em Phys. Rev. Lett.}, 121(11):111302, 2018, 1805.12562.

\bibitem{Sirunyan:2019twz}
Albert~M Sirunyan et~al.
\newblock {Measurements of the Higgs boson width and anomalous $HVV$ couplings
  from on-shell and off-shell production in the four-lepton final state}.
\newblock {\em Phys. Rev. D}, 99(11):112003, 2019, 1901.00174.

\bibitem{ATLAS-CONF-2020-052}
{Combination of searches for invisible Higgs boson decays with the ATLAS
  experiment}.
\newblock Technical Report ATLAS-CONF-2020-052, CERN, Geneva, Oct 2020.

\bibitem{Lundstrom:2008ai}
Erik Lundstrom, Michael Gustafsson, and Joakim Edsjo.
\newblock {The Inert Doublet Model and LEP II Limits}.
\newblock {\em Phys. Rev.}, D79:035013, 2009, 0810.3924.

\bibitem{Akerib:2013tjd}
D.~S. Akerib et~al.
\newblock {First results from the LUX dark matter experiment at the Sanford
  Underground Research Facility}.
\newblock {\em Phys. Rev. Lett.}, 112:091303, 2014, 1310.8214.

\bibitem{Kalinowski:2018kdn}
Jan Kalinowski, Wojciech Kotlarski, Tania Robens, Dorota Sokolowska, and
  Aleksander~Filip Zarnecki.
\newblock {Exploring Inert Scalars at CLIC}.
\newblock {\em JHEP}, 07:053, 2019, 1811.06952.

\bibitem{deBlas:2018mhx}
R.~Franceschini et~al.
\newblock {The CLIC Potential for New Physics}.
\newblock 2018, 1812.02093.

\bibitem{Moretti:2001zz}
Mauro Moretti, Thorsten Ohl, and Jurgen Reuter.
\newblock {O'Mega: An Optimizing matrix element generator}.
\newblock pages 1981--2009, 2001, hep-ph/0102195.

\bibitem{Kilian:2007gr}
Wolfgang Kilian, Thorsten Ohl, and Jurgen Reuter.
\newblock {WHIZARD: Simulating Multi-Particle Processes at LHC and ILC}.
\newblock {\em Eur. Phys. J.}, C71:1742, 2011, 0708.4233.

\bibitem{Staub:2015kfa}
Florian Staub.
\newblock {Exploring new models in all detail with SARAH}.
\newblock {\em Adv. High Energy Phys.}, 2015:840780, 2015, 1503.04200.

\bibitem{Porod:2003um}
Werner Porod.
\newblock {SPheno, a program for calculating supersymmetric spectra, SUSY
  particle decays and SUSY particle production at e+ e- colliders}.
\newblock {\em Comput. Phys. Commun.}, 153:275--315, 2003, hep-ph/0301101.

\bibitem{Porod:2011nf}
W.~Porod and F.~Staub.
\newblock {SPheno 3.1: Extensions including flavour, CP-phases and models
  beyond the MSSM}.
\newblock {\em Comput. Phys. Commun.}, 183:2458--2469, 2012, 1104.1573.

\bibitem{Linssen:2012hp}
Lucie Linssen, Akiya Miyamoto, Marcel Stanitzki, and Harry Weerts.
\newblock {Physics and Detectors at CLIC: CLIC Conceptual Design Report}.
\newblock 2012, 1202.5940.

\bibitem{Hocker:2007ht}
Andreas Hocker et~al.
\newblock {TMVA - Toolkit for Multivariate Data Analysis}.
\newblock 2007, physics/0703039.

\bibitem{Sokolowska:2019xhe}
Dorota Sokolowska, Jan Kalinowski, Jan Klamka, Pawel Sopicki, Aleksander~Filip
  Zarnecki, Wojciech Kotlarski, and Tania Robens.
\newblock {Inert Doublet Model signatures at future $e^+e^-$ colliders}.
\newblock {\em PoS}, EPS-HEP2019:570, 2020, 1911.06254.

\bibitem{Zarnecki:2020swm}
Aleksander~Filip Zarnecki, Jan Kalinowski, Jan Klamka, Pawel Sopicki, Wojciech
  Kotlarski, Tania Robens, and Dorota Sokolowska.
\newblock {Searching Inert Scalars at Future e$^+$e$^-$ Colliders}.
\newblock In {\em {International Workshop on Future Linear Colliders (LCWS
  2019) Sendai, Miyagi, Japan, October 28-November 1, 2019}}, 2020, 2002.11716.

\bibitem{Zarnecki:2020nnw}
Aleksander~Filip Zarnecki, Jan Kalinowski, Jan Klamka, Pawel Sopicki, Wojciech
  Kotlarski, Tania~Natalie Robens, and Dorota Sokolowska.
\newblock {Searching inert scalars at future $e^+e^-$ colliders}.
\newblock {\em PoS}, CORFU2019:047, 2020.

\bibitem{Klamka:2728552}
Jan~Franciszek Klamka.
\newblock {Searching for Inert Doublet Model scalars at high energy CLIC}.
\newblock CERN-THESIS-2020-098, 2020.

\bibitem{ATL-PHYS-PUB-2019-005}
{Expected performance of the ATLAS detector at the High-Luminosity LHC}.
\newblock Technical Report ATL-PHYS-PUB-2019-005, CERN, Geneva, Jan 2019.

\bibitem{Collaboration:2650976}
The~CMS Collaboration.
\newblock {Expected performance of the physics objects with the upgraded CMS
  detector at the HL-LHC}.
\newblock Technical Report CMS-NOTE-2018-006. CERN-CMS-NOTE-2018-006, CERN,
  Geneva, Dec 2018.

\bibitem{Abada:2019ono}
A.~Abada et~al.
\newblock {HE-LHC: The High-Energy Large Hadron Collider}.
\newblock {\em Eur. Phys. J. ST}, 228(5):1109--1382, 2019.

\bibitem{Benedikt:2018csr}
A.~Abada et~al.
\newblock {FCC-hh: The Hadron Collider}.
\newblock {\em Eur. Phys. J. ST}, 228(4):755--1107, 2019.

\bibitem{Delahaye:2019omf}
Jean~Pierre Delahaye, Marcella Diemoz, Ken Long, Bruno Mansoulié, Nadia
  Pastrone, Lenny Rivkin, Daniel Schulte, Alexander Skrinsky, and Andrea
  Wulzer.
\newblock {Muon Colliders}.
\newblock 2019, 1901.06150.

\bibitem{Robens:2019kga}
Tania Robens, Tim Stefaniak, and Jonas Wittbrodt.
\newblock {Two-real-scalar-singlet extension of the SM: LHC phenomenology and
  benchmark scenarios}.
\newblock {\em Eur. Phys. J.}, C80(2):151, 2020, 1908.08554.

\bibitem{Papaefstathiou:2020lyp}
Andreas Papaefstathiou, Tania Robens, and Gilberto Tetlalmatzi-Xolocotzi.
\newblock {Triple Higgs Boson Production at the Large Hadron Collider with Two
  Real Singlet Scalars}.
\newblock 2020, 2101.00037.

\bibitem{Coimbra:2013qq}
Rita Coimbra, Marco O.~P. Sampaio, and Rui Santos.
\newblock {ScannerS: Constraining the phase diagram of a complex scalar singlet
  at the LHC}.
\newblock {\em Eur. Phys. J.}, C73:2428, 2013, 1301.2599.

\bibitem{Costa:2015llh}
Raul Costa, Margarete Muehlleitner, Marco O.~P. Sampaio, and Rui Santos.
\newblock {Singlet Extensions of the Standard Model at LHC Run 2: Benchmarks
  and Comparison with the NMSSM}.
\newblock {\em JHEP}, 06:034, 2016, 1512.05355.

\bibitem{Muhlleitner:2020wwk}
Margarete Mühlleitner, Marco O.~P. Sampaio, Rui Santos, and Jonas Wittbrodt.
\newblock {ScannerS: Parameter Scans in Extended Scalar Sectors}.
\newblock 2020, 2007.02985.

\bibitem{Heinemeyer:2013tqa}
J~R Andersen et~al.
\newblock {Handbook of LHC Higgs Cross Sections: 3. Higgs Properties}.
\newblock 2013, 1307.1347.

\bibitem{Alwall:2014hca}
J.~Alwall, R.~Frederix, S.~Frixione, V.~Hirschi, F.~Maltoni, O.~Mattelaer,
  H.~S. Shao, T.~Stelzer, P.~Torrielli, and M.~Zaro.
\newblock {The automated computation of tree-level and next-to-leading order
  differential cross sections, and their matching to parton shower
  simulations}.
\newblock {\em JHEP}, 07:079, 2014, 1405.0301.

\bibitem{Hirschi:2015iia}
Valentin Hirschi and Olivier Mattelaer.
\newblock {Automated event generation for loop-induced processes}.
\newblock {\em JHEP}, 10:146, 2015, 1507.00020.

\bibitem{Bahr:2008pv}
M.~Bahr et~al.
\newblock {Herwig++ Physics and Manual}.
\newblock {\em Eur. Phys. J.}, C58:639--707, 2008, 0803.0883.

\bibitem{Gieseke:2011na}
S.~Gieseke et~al.
\newblock {Herwig++ 2.5 Release Note}.
\newblock 2011, 1102.1672.

\bibitem{Arnold:2012fq}
K.~Arnold et~al.
\newblock {Herwig++ 2.6 Release Note}.
\newblock 2012, 1205.4902.

\bibitem{Bellm:2013hwb}
J.~Bellm et~al.
\newblock {Herwig++ 2.7 Release Note}.
\newblock 2013, 1310.6877.

\bibitem{Bellm:2015jjp}
Johannes Bellm et~al.
\newblock {Herwig 7.0/Herwig++ 3.0 release note}.
\newblock {\em Eur. Phys. J.}, C76(4):196, 2016, 1512.01178.

\bibitem{Bellm:2017bvx}
Johannes Bellm et~al.
\newblock {Herwig 7.1 Release Note}.
\newblock 2017, 1705.06919.

\bibitem{Bellm:2019zci}
Johannes Bellm et~al.
\newblock {Herwig 7.2 release note}.
\newblock {\em Eur. Phys. J. C}, 80(5):452, 2020, 1912.06509.

\bibitem{Aaboud:2017rel}
M.~Aaboud et~al.
\newblock {Search for heavy ZZ resonances in the $\ell ^+\ell^-\ell ^+\ell ^-$
  and $\ell ^+\ell ^-\nu \bar{\nu }$ final states using proton–proton
  collisions at $\sqrt{s}= 13$ $\text {TeV}$ with the ATLAS detector}.
\newblock {\em Eur. Phys. J.}, C78(4):293, 2018, 1712.06386.

\bibitem{Sirunyan:2018qlb}
Albert~M Sirunyan et~al.
\newblock {Search for a new scalar resonance decaying to a pair of Z bosons in
  proton-proton collisions at $\sqrt{s}=13 $ TeV}.
\newblock {\em JHEP}, 06:127, 2018, 1804.01939.
\newblock [Erratum: JHEP03,128(2019)].

\bibitem{Aaboud:2018bun}
Morad Aaboud et~al.
\newblock {Combination of searches for heavy resonances decaying into bosonic
  and leptonic final states using 36 fb$^{-1}$ of proton-proton collision data
  at $\sqrt{s} = 13$ TeV with the ATLAS detector}.
\newblock {\em Phys. Rev.}, D98(5):052008, 2018, 1808.02380.

\bibitem{Sirunyan:2018two}
Albert~M Sirunyan et~al.
\newblock {Combination of searches for Higgs boson pair production in
  proton-proton collisions at $\sqrt{s} = $ 13 TeV}.
\newblock {\em Phys. Rev. Lett.}, 122(12):121803, 2019, 1811.09689.

\bibitem{Aad:2019uzh}
Georges Aad et~al.
\newblock {Combination of searches for Higgs boson pairs in $pp$ collisions at
  $\sqrt{s} = $13 TeV with the ATLAS detector}.
\newblock {\em Phys. Lett.}, B800:135103, 2020, 1906.02025.

\bibitem{Cepeda:2019klc}
M.~Cepeda et~al.
\newblock {Report from Working Group 2}.
\newblock {\em CERN Yellow Rep. Monogr.}, 7:221--584, 2019, 1902.00134.

\bibitem{Aaboud:2017gsl}
Morad Aaboud et~al.
\newblock {Search for heavy resonances decaying into $WW$ in the $e\nu\mu\nu$
  final state in $pp$ collisions at $\sqrt{s}=13$ TeV with the ATLAS detector}.
\newblock {\em Eur. Phys. J.}, C78(1):24, 2018, 1710.01123.

\bibitem{ATL-PHYS-PUB-2018-022}
{HL-LHC prospects for diboson resonance searches and electroweak vector boson
  scattering in the $WW/WZ\to\ell\nu qq$ final state}.
\newblock Technical Report ATL-PHYS-PUB-2018-022, CERN, Geneva, Oct 2018.

\bibitem{Papaefstathiou:2020iag}
Andreas Papaefstathiou and Graham White.
\newblock {The Electro-Weak Phase Transition at Colliders: Confronting
  Theoretical Uncertainties and Complementary Channels}.
\newblock 10 2020, 2010.00597.

\end{thebibliography}

\end{document}